\documentclass[12pt]{article}
\usepackage{graphicx}
\usepackage{dcolumn}
\usepackage{bm}
\usepackage{amsfonts}
\usepackage{amsthm}
\usepackage{amsmath}
\usepackage{amssymb}
\usepackage{epsfig}
\usepackage{color}
\usepackage{textcomp}

\def\be{\begin{equation}}
\def\ee{\end{equation}}
\def\ba{\begin{eqnarray}}
\def\ea{\end{eqnarray}}

\def\bdm{\begin{displaymath}}
\def\edm{\end{displaymath}}
\def\la{~\mbox{\raisebox{-.6ex}{$\stackrel{<}{\sim}$}}~}
\def\ga{~\mbox{\raisebox{-.6ex}{$\stackrel{>}{\sim}$}}~}
\def\bq{\begin{quote}}
\def\eq{\end{quote}}

 at 10truept

\newcommand{\bea}{\begin{eqnarray}}
\newcommand{\eea}{\end{eqnarray}}

\newcommand{\bi}{\begin{itemize}}
\newcommand{\ei}{\end{itemize}}

\newcommand{\beq}{\begin{equation}}
\newcommand{\eeq}{\end{equation}}
\newcommand{\beqa}{\begin{eqnarray}}
\newcommand{\eeqa}{\end{eqnarray}}

\newcommand{\del}{\partial}

\def\la{~\mbox{\raisebox{-.6ex}{$\stackrel{<}{\sim}$}}~}
\def\ga{~\mbox{\raisebox{-.6ex}{$\stackrel{>}{\sim}$}}~}

\def\ltap{\ \raise.3ex\hbox{$<$\kern-.75em\lower1ex\hbox{$\sim$}}\ }
\def\gtap{\ \raise.3ex\hbox{$>$\kern-.75em\lower1ex\hbox{$\sim$}}\ }
\def\gl{\ \raise.5ex\hbox{$>$}\kern-.8em\lower.5ex\hbox{$<$}\ }
\def\roughly#1{\raise.3ex\hbox{$#1$\kern-.75em\lower1ex\hbox{$\sim$}}}

\begin{document}

\thispagestyle{empty}
\begin{flushright}
May 2014
\end{flushright}
\vspace*{.9cm}
\begin{center}
{\Large \bf Vacuum Energy Sequestering: The Framework and}\\
\vskip.3cm
{\Large \bf Its Cosmological Consequences}\\

\vspace*{1cm} {\large Nemanja Kaloper$^{a, }$\footnote{\tt
kaloper@physics.ucdavis.edu} and Antonio Padilla$^{b, }$\footnote{\tt
antonio.padilla@nottingham.ac.uk} }\\
\vspace{.5cm} {\em $^a$Department of Physics, University of
California, Davis, CA 95616, USA}\\
\vspace{.5cm} {\em $^b$School of Physics and Astronomy, 
University of Nottingham, Nottingham NG7 2RD, UK}\\

\vspace{1.5cm} ABSTRACT
\end{center}
Recently we suggested a reformulation of General Relativity which completely sequesters from gravity {\it all} of the vacuum energy 
from a protected matter sector, assumed to contain the Standard Model.  Here we elaborate further on the mechanism, presenting additional details of how it cancels all loop corrections and renders all contributions from phase transitions automatically small. We also consider cosmological consequences in more detail and show that the mechanism is consistent with a variety of inflationary models that make a universe big and old. 
We discuss in detail the underlying assumptions behind the dynamics of our  proposal, and elaborate on the relationship of the physical interpretation of divergent operators in quantum field theory and the apparent `acausality' which our mechanism seems to entail, which we argue is completely harmless. It is merely a reflection of the fact that
any UV sensitive quantity in quantum field theory cannot be calculated from first principles, but is an input whose numerical value must be measured. 
We also note that since the universe should be compact in spacetime, and so will collapse in the future, the current phase of acceleration with $w_{DE}\approx-1$ is just a transient. This could be tested by future cosmological observations.

\vfill \setcounter{page}{0} \setcounter{footnote}{0}
\newpage

\section{Introduction}

Cosmological observations suggest that the cosmological constant is different from zero. Yet we have no clear and compelling argument which would explain the observed scale of the cosmological constant in quantum field theory (QFT). This fact is recognized under the term ``the cosmological constant problem'' \cite{zeldovich,wilczek,wein}.  In a nutshell, the problem arises because of the universality of gravity. 
The Equivalence Principle of General Relativity (GR), which controls how energy distorts geometry, 
posits that all forms of energy curve space-time. Since in QFT even the vacuum possesses energy density, given by the resummation of the QFT bubble diagrams, this means that the vacuum geometry generically must be curved. The scale of the curvature is 
$L_{vac} \sim (\sqrt{G_N \rho_{vac}})^{-1}$. Cosmological observations then constrain it to be $L_{vac} \ga 10^{24}$ cm, implying that the energy density of the vacuum must satisfy $\rho_{vac} \la (10^{-3} {\rm eV})^4$. However, attempts to estimate the contributions to $\rho_{vac}$  in
QFT, using the available scales in Nature, exceed this value manyfold. An early example of this contradiction led Pauli to famously quip 
``that the radius of the world would not even reach to the Moon'' (cf. $3.8 \times 10^7$ cm) after he had estimated the vacuum energy contributions down to the scale set by the classical radius of the electron \cite{ruzi}. Estimates involving higher energy cutoffs yield energy densities of the vacuum far in excess of this value, at least as high as $(\rm{TeV})^4$, and possibly as high as Planckian scales, $M_{Pl}^4$. 

The structure of GR, namely the underlying diffeomorphism invariance, allows one to freely add a classical contribution to the cosmological constant
and tune it with tremendous precision to cancel  off the vacuum energy. This means one needs to pick the classical piece to be the opposite of whatever the (regulated!) result of a field theory calculation of the quantum vacuum energy yields, plus an extra $(10^{-3} {\rm eV})^4$, once one chooses to satisfy the 
cosmological observations with the leftover remaining after the cancellation. However, this choice is unstable in any perturbative scheme for the computation of vacuum energy: any change of the matter sector 
parameters or inclusion of higher order loop corrections to the vacuum energy shifts its value significantly, often by ${\cal O}(1)$ in the units of the 
UV cutoff\footnote{If exact, SUSY and/or conformal symmetry can enforce the vanishing of vacuum
energy. There are some cases where the corrections might be smaller, of quadratic order in the cutoff \cite{akhmedov,ctco,rattazi}. However the corrections are quartic in the cutoff in generic cases, with interactions; 
without supersymmetry and conformal symmetry, the vacuum energy is given 
by the fourth power of the breaking scale of these symmetries\cite{wein, dreitlein,linde,veltman,sirlin}.}.
One must then retune the classical term by hand order by order in perturbation theory. This is discomforting. 
The classical contribution to the cosmological constant is a separately conserved quantity, coded in the initial conditions of the universe. So the required retuning of its value needed to cancel the subsequent corrections the vacuum energy in quantum theory means that one - in reality - must choose a completely different approximate description of the cosmic history to attain the required smallness of the vacuum curvature at late times. 

One could try to argue how this might be a red herring, by proposing to sum all the quantum contributions to the total vacuum energy right away. After that one could simply pick the required value of the classical contribution, however tuned, and be done with it. One might even wonder if the large corrections from known field theory degrees of freedom are merely a problem of perturbation theory rather than  a real 
physical issue\footnote{Implausible as this might sound, given that no convincing alternatives to perturbation theory calculations of cosmological constant have been offered to date.}.
However, this is clearly impossible if one does not know the full QFT up to whatever the fundamental cutoff may be. One can merely guess the quantum contents of the universe beyond the TeV scale, and so cannot be sure just what it is that needs to be cancelled. Further, quantum corrections to the vacuum energy coming from phase transitions at late times would not be properly accounted for either.

In fact, the subtleties are brought into focus when one recounts how one really deals with UV-divergent quantities. In QFT they must be renormalized: after a suitable regularization, which at a technical level introduces a cutoff (that renders the divergence formally finite, but very sensitive to whatever lies beyond the cutoff) the divergence -- ie the contribution which diverges as the cutoff is sent to infinity -- must be subtracted by the bare counterterm. This is the real role of the classical contribution to the cosmological constant: it is the bare counterterm which one must keep in the theory for the purpose of renormalization. This is true even in effective field theories with a hard cutoff \cite{polchinski}, where one still renormalizes away the cutoff-dependent terms which signify the dependence on the unknown short distance physics. The remainder is finite. However since it depends on an arbitrary subtraction scale it {\it cannot} be computed from first principles. Thus since the vacuum energy is divergent its final value cannot be predicted within QFT. It must be {\it measured}, just as, for example, a quadratically divergent mass of a scalar field. Once the measurement is performed, and the terminal numerical value obtained, predictions can be made for other observables, which depend on the renormalized vacuum energy, but are not divergent themselves. 

Now, the measurement of the cosmological constant may seem to be a simple exercise, since, after all, it's just one number. However, this depends on the nature of the mechanism of subtracting the UV-sensitive part from it. 
If the subtraction scheme involves local fields, then the leftover value could in principle vary in spacetime and needs to be measured more precisely. In other words, once must be able to distinguish between the cosmological constant, and the contributions which vary in the far IR very slowly, having weak dependence on very low momenta.
The problem with measuring the vacuum energy is that it is a quantity which characterizes an object of co-dimension zero. To measure it, one therefore ultimately needs a detector of the same co-dimension: namely, the whole universe.  Only in this way can one assure that the remainder is really a constant, being the same everywhere and at all times. The measurement with an arbitrary precision thus must be non-local, in space and in time, from the viewpoint of an observer of a smaller co-dimension cohabiting the universe. Remarks about the necessity of a nonlocal measurement of cosmological constant were also noted in \cite{degrav}. 

Thus asking why the cosmological constant is small, large, or anything in between ultimately 
does not -- really -- make sense in the context of QFT coupled of gravity. To see what it is, one must regulate it, renormalize it and {\it measure} it. The one aspect of the problem which remains, however, is why is the evaluation of the leftover cosmological constant radiatively unstable? Or in other words, why do the measurements of the cosmological constant after regularization require large modifications of the finite subtractions between consecutive orders in perturbation theory? This issue is fully analogous to the gauge hierarchy problem in QFT, and is well defined in QFT coupled to gravity. Yet there are precious few clues as to how to address it, in contrast to various extensions of the Standard Model beyond the electroweak breaking scale designed precisely to address the gauge hierarchy problem\footnote{The absence of direct signatures from the LHC of any such new BSM physics to date may raise concerns whether naturalness is realized in Nature. In what follows, however, we shall ignore this.}. Note, that this applies to {\it any} attempt to address the cosmological constant problem within the realm of local QFT. 

At first sight, not only does this perspective invoke nonlocality, but it also resembles a `circular argument'. There is a precedent for this logic, however. 
Recall how a notion of force is defined in Newtonian mechanics. To formulate the second law, $ m \vec a = \vec F$, one needs to first introduce a standard of mass whose accelerations in response to various externally applied agents calibrate their forces. Only after such a calibration has been completed, can one begin to make predictions for all other masses in a Newtonian universe. The mass standard must be taken out of the clockwork, since no predictions can be made for it: its motion defines the forces, rather than the other way around. Of course, since the mass characterizes objects whose co-dimension is $3$, the sector of the universe whose evolution cannot be predicted is a mere worldline, a set of measure zero. This is far easier to ignore. That
the mass scales in QFT of the Standard Model are all set by the Higgs, whose own mass is quadratically divergent, and therefore can only be determined {\it a posteriori}, by measurement, only reaffirms this view further.
The challenge, therefore, is not to ask why the cosmological constant is small, but why it is radiatively stable.

Motivated by this `vacuum' of natural protection mechanisms of the vacuum energy, in a recent letter \cite{kalpad1} we proposed a mechanism that ensures that {\it all} vacuum energy contributions from a protected matter sector are in fact sequestered from (semi)classical gravity. Further the mechanism automatically renders all the contributions to  vacuum energy coming from the phase transitions small, and hence observationally harmless. Crucial for the mechanism is that we only consider the theory in the decoupling limit of gravity, prohibiting gravitons as internal lines of the vacuum energy corrections\footnote{This limit of the problem is in fact precisely how Zeldovitch formulated it in \cite{zeldovich}.}.
In this paper we will expand on that proposal, providing further details of the mechanism and its consequences. The sequestering works at each and every order in perturbation theory, so there is no need to retune the classical cosmological constant when higher loop corrections are included. Instead, as we noted above, one is left with a residual cosmological constant, which is automatically radiatively stable, being completely independent of the vacuum energy contributions from the protected sector.  In a universe that grows old and big, the residual cosmological constant is naturally small, along with any contributions from phase transitions in the early universe. To be clear, our mechanism takes care of all vacuum energy contributions from a protected matter sector, which we take to include the Standard Model of Particle Physics, but has nothing to say about virtual graviton loops\footnote{The graviton contributions to the vacuum energy are context dependent (see eg. 
\cite{shamiteva} for an exploration of loop contributions to vacuum energy in string theory). The Standard Model contributions are not, once we assume  QFT to describe them. The result of our protection mechanism might be de-sensitized from gravity, for example, if we can supersymmetrize gravitational sector down to a 
millimeter scale, in which case gravitational contributions to vacuum energy in field theory would never exceed the observational bounds. While this looks interesting, further work is needed to determine if such an extension could work.}.

Our mechanism is based on the introduction of global constraints in the formulation of the action describing the matter coupled to gravity. We postulate two such constraints, by promoting the classical contribution to the cosmological constant, $\Lambda$, and the dimensionless parameter which controls the protected sector scales relative to a fixed Planck scale, $\lambda \propto m_{phys}/M_{Pl}$, to global variables in the variational principle,
and extend the action by adding to it a term $\sigma(\Lambda, \lambda)$ {\it outside} of the intergal. Then diffeomorphism invariance guarantees that the vacuum energy always scales with $\lambda$ in the same way, regardless of the order in the loop expansion, and the constraints imposed by the variation with respect to the global variables amount to dynamically determining the value of $\Lambda$ which always precisely cancels the vacuum energy. 
Since the theory remains completely diffeormorphism invariant and locally Poincar\'e invariant, no new local degrees of freedom appear. So locally the theory looks just like a usual QFT coupled to gravity, but with an
 {\it aposteriori} cosmological constant determined by a non-local measurement intrinsic to the universe's dynamics - as it should be. An alternative way of describing the mechanism is to integrate out the auxiliary variables $\Lambda$ and $\lambda$. Then our mechanism can be understood simply as setting all scales in the protected matter sector to be functionals of the $4$-volume element of the universe\footnote{An equivalent interpretation is to fix the protected field theory scales, and make the Planck scale a functional of the world volume of the universe. 
}. This ensures that the vacuum energy quantum corrections consistently drop out order by order in perturbation theory, yet the local dynamics remains the same as in the standard approach. 
Furthermore, the mechanism is consistent with phenomenological requirements, specifically with large hierarchies between the Planck scale,
electroweak scale and vacuum curvature scale, and with early universe cosmology including inflation.
In the Letter \cite{kalpad1} we have illustrated the last point by showing that the mechanism is completely harmonious with the Starobinsky inflation. Here, in contrast we will show that it can also coexist with a model where the last 60 efolds of inflation are driven by a quadratic potential from the protected sector.
Nevertheless, the global cosmology must differ. For the protected matter scales to be nonzero, the universe should have a finite space-time volume, being spatially compact and crunching in the future. This represents the main prediction of our theory to date. An immediate corollary is that the current phase of acceleration cannot last forever, which means that $w_{DE} \approx -1$ is merely a transient phase. Ergo, the dark energy cannot be everpresent cosmological constant\footnote{Although it can approximate a cosmological constant very accurately for a long period.}. In addition, since the cosmological contributions of local events are weighed down by the large spacetime volume of the universe, the contributions to vacuum energy from phase transitions are automatically small. Hence in our framework the residual net cosmological constant, which   sources the curvature of the vacuum, is 
\begin{itemize}
\item purely classical, set by the complete evolution of the geometry;
\item a `cosmic average' of the values of non-constant sources;
\item  automatically small in universes which grow old and big. 
\end{itemize}

This paper is organized as follows. In section \ref{sec:ccp}, we review the cosmological constant problem, outlining the key features, and ending with a discussion of Weinberg's venerable {\it no-go} theorem prohibiting a local field theory adjustment mechanism \cite{wein}. In section \ref{prop} we present our main proposal in detail, explaining how the cancellation of vacuum energy works, focusing on a pair of symmetries that are ultimately responsible for this cancellation. In section \ref{sec:hist} we study the kinematics of our theory in some detail. In particular we estimate the relevant historic cosmic averages, demonstrating that the residual cosmological constant will not exceed the critical density today, and show that the 
contributions from (many!) astrophysical black holes in section \ref{sec:bh} are small. 
In section   \ref{sec:ex} we outline the procedure for finding general cosmological solutions which satisfy the global constraints by their selection from the families of known solutions of General Relativity. The very important question of  contributions to vacuum energy from early universe phase transitions  are the topic of section \ref{sec:phase}, where we show that after the transition they are automatically small in large old universes. In section \ref{sec:masses}, we show how our mechanism is compatible with the observed mass scales in particle physics, and how this can be achieved without introducing any new hierarchies.
Generalizations of our proposal are presented in section \ref{sec:gen}, including the effect of radiative corrections to the Planck scale. 
 Section \ref{sec:inflation} is devoted to inflation. Commenting on consistency with Starobinsky inflation, we also show that single monomial inflationary models, and specifically quadratic inflation, driven by fields from the protected sector, are also compatible with our mechanism. We further comment on the possible conflict with eternal inflation, which generically yields universes with infinite volume. Finally, in section \ref{sec:conc}, we briefly summarize some key features and observational signatures of our proposal.

\section{The cosmological constant problem} \label{sec:ccp}

We begin with a brief review of the vacuum energy problem from the viewpoint of QFT coupled to gravity. Further details can be found in many reviews of the subject, for example \cite{wein,Yokoyama,Dolgov, Martin,cliff}. So: consider a QFT of matter, which -- to be complete in the flat space limit -- requires a specification of a UV regulator. By a UV regulator, we mean a procedure for rendering the divergences in renormalizable QFTs formally finite, so that they can be systematically subtracted by the addition of bare counterterms.  This could just be a hard cutoff 
in nonrenormalizable theories, which is really an avatar of the full renormalization procedure in the UV completion of the theory (assuming that one exists). Couple this theory 
to a covariant semi-classical theory of gravity universally (i.e. minimally), by defining the source of gravity to be stress-energy of the matter QFT. Then,
by the Equivalence Principle, the vacuum energy of the QFT, which corresponds to the resummation of the bubble diagrams in the loop expansion and drops out from the expressions for the scattering amplitudes in flat space\footnote{Which is guaranteed by the fact that in flat space field theory one can freely recalibrate the zero point energies of any matter fields.} couples to gravity via the covariant measure, $-V_{vac} \int d^4 x \sqrt{g}$, where $g=|\det g_{\mu\nu}|$ is the determinant of the metric $g_{\mu\nu}$. Hence the vacuum energy sources gravity, yielding an energy-momentum tensor $T_{\mu\nu}=-V_{vac} g_{\mu\nu}$. Since $V_{vac}$ is divergent, we should also include a bare cosmological constant -- i.e. the `classical contribution' -- as a counter-term in the action, $-\Lambda_{bare}\int d^4 x \sqrt{g}$, so that it is actually the combination  $\Lambda_{total}=\Lambda_{bare}+V_{vac}$ that gravitates. Cosmology then requires that this combination should not exceed the observed dark energy scale, i.e., $\Lambda_{total}=\Lambda_{bare}+V_{vac} \la (10^{-3} {\rm eV})^4$.

In  practice, one computes $V_{vac}$ in perturbation theory, using the loop expansion in flat 
space\footnote{Note, that the results are consistent with the UV-sensitive terms computed in curved space using covariant regularization techniques \cite{myers,soloduk}.}. This means, one truncates the infinite series of bubble diagram contributions to a desired precision in the loop expansion, evaluates these diagrams and sums them up. Eg. at, say, one loop, one simply evaluates 
the vacuum energy to one loop, $V_{vac}=V_{vac}^\text{tree}+V_{vac}^\text{1-loop}$, and then subtracts the infinity in $V_{vac}$ by 
\begin{itemize}
\item regulating $V_{vac}$ with, eg. a hard cutoff;
\item adding the bare term 
$\Lambda_{bare}$ to subtract away the UV sensitive contribution from $V_{vac}$;
\item  tuning the remainder in $\Lambda_{bare}$ so that the observational constraint is satisfied.
\end{itemize} 
Although the definition of the calculated $\Lambda_{total} = \Lambda_{bare} + V_{vac}$ ensures that it is a constant, for it to actually be measured one {\it must} continue measuring it across the whole worldvolume of the universe, as we explained in the introduction. This fact is obscured by the description of the flat space subtraction procedure utilized here -- since it is commonplace -- because the subtraction procedure we employ is local (being imposed by an external observer who is computing $\Lambda_{total}$), and so the residual leftover $\Lambda_{total}$ is a constant by consequence.

Note that this implies that $\Lambda_{total}$ can be very different from either $\Lambda_{bare}$ or $V_{vac}$.
It is not predicted - it is to be measured, as we stressed above, and so it can be anything. This is standard in renormalization in QFT. So, saying that $\Lambda_{total} \ll V_{vac}$ is not a problem by itself. 
The problem is that the renormalization procedure sketched above is {\it not} perturbatively stable. Eg, generically, the regulated two loop correction is $V_{vac}^\text{2-loop} \sim V_{vac}^\text{1-loop}$, which is much greater than $\Lambda_{total}$ that remains after the subtraction of the UV-sensitive contribution at 1-loop level. This means that the bare (`classical') cosmological constant $\Lambda_{bare}$, which was selected with a high degree of precision to cancel the UV-sensitive contribution to the one loop vacuum energy needs to be retuned by ${\cal O}(1)$ in the units of the UV cutoff, 
in order to yield a new two-loop $\Lambda_{total}$ that remains compatible with the observational bound. 
Once gravity is turned on, so that we can view $\Lambda_{bare}$ as a classical conserved quantity fixed by some cosmological initial condition, this implies that the initial condition must be altered dramatically in order to readjust the two-loop value of $\Lambda_{total}$ to satisfy the bound. Thus the observable value of the total cosmological constant is not a reliable, stable prediction of the theory. It is very sensitive to both the details of the UV physics, which are unknown, and to the cosmological initial conditions, which are notoriously difficult to reconstruct. This issue repeats at each successive order in the loop expansion, indicating the radiative instability of the small value of the observable $\Lambda_{total}$. 
 
A simple illustrative example  is provided by a massive scalar field theory with quartic self-coupling and minimal coupling to gravity. At one loop, the relevant Feynman diagrams correspond to a single scalar loop with  external graviton legs carrying zero momenta.  The sum of all such diagrams generates a term $-V^{\phi, \text{1-loop}}_{vac} \int d^4 x \sqrt{g}$, so it suffices to calculate just one. Perhaps the simplest contribution is the tadpole diagram, given by \cite{Martin}
\be
\text{tadpole}= \frac12\int \frac{d^4 k}{(2\pi)^4}\left[\frac{k^\mu k^\nu-\frac12  \eta^{\mu\nu}\ (k^2+m^2)}{k^2+m^2}\right]=-\frac{i}{2} \eta^{\mu\nu} V^{\phi, \text{1-loop}}_{vac} \, .
\ee
Using dimensional regularization, one finds \cite{Martin}
\be
V^{\phi, \text{1-loop}}_{vac}= -\frac{m^4} {(8 \pi)^2}\left[\frac{2}{\epsilon}+\text{finite}+\ln\left(\frac{M_{UV}^2} {m^2}\right)\right] \, ,
\ee
where $M_{UV}$ is the UV regulator scale. The bare counterterm one would add to cancel the divergence would therefore be
\be
\Lambda_{bare} = \frac{m^4} {(8 \pi)^2}\left[\frac{2}{\epsilon} + \ln\left(\frac{M_{UV}^2}{{\cal M}^2}\right)\right] \, ,
\ee
where ${\cal M}$ is the subtraction point. The 1-loop renormalized cosmological constant would then be
\be
\Lambda_{ren} = V^{\phi, \text{1-loop}}_{vac}+\Lambda_{bare} = \frac{m^4} {(8 \pi)^2}
\left[\ln\left(\frac{m^2}{{\cal M}^2} \right) - \text{finite} \right] \, .
\ee
Note that the remainder depends on $\ln{\cal M}$, illustrating the dependence of the renormalized cosmological constant on the arbitrary subtraction scale ${\cal M}$. This is why the value of $\Lambda_{ren}$ can only be fixed by a measurement.
At two loops, we consider the so-called  scalar ``figure of eight"  with external graviton legs. Its contribution to vacuum energy is given by $V^{\phi, \text{2-loop}}_{vac} \sim \lambda m^4$. For perturbative theories without finely tuned couplings, where $\lambda \sim {\cal O}(0.1)$, (as for example the Standard Model Higgs) the higher loop corrections remain competitive with the leading order contributions.

One could try to improve the situation by designing dynamical mechanisms which cancel the vacuum energy contributions to the  total cosmological constant order by order in the loop expansion. An example is provided by either supersymmetric theories or conformal field theories. In both cases, it is the underlying unbroken symmetry which automatically sets the vacuum energy at any order in the loop expansion to zero. In the case of supersymmetry, this follows from the cancellations of loop diagrams between bosons and fermions with degenerate masses. For conformal theories, it is the unbroken conformal symmetry which simply scales away all the dimensional characteristics of the vacuum. Both examples represent technically natural solutions of the problem: the vacuum energy, and the total cosmological constant vanish as a consequence of the underlying symmetry. However, in the real world, both of these symmetries, if they exist, are broken, at least below the electroweak breaking scale. The prediction which follows from the breaking, and which is in fact technically natural, is that the resulting vacuum energy should be at least $M_{EW}^4$. Restoring either supersymmetry or conformal symmetry (or both) would render the vacuum energy to be zero, and so these symmetries could be protecting the cosmological constant from the corrections larger than $M_{EW}^4$. However, in our world the observed value of the cosmological constant must be much smaller, by at least 60 orders of magnitude. 

An alternative approach could be to look for some dynamical extension of the minimal framework of QFT coupled to gravity, where a symmetry protecting a small total cosmological constant could be hidden. This was an idea behind many past attempts to address the cosmological constant problem via the adjustment 
mechanism. This approach is obstructed by the venerable Weinberg's no-go theorem \cite{wein}, which precludes a dynamical adjustment mechanism of the cosmological constant in the framework of an effective QFT coupled to gravity. We review it here, as it provides a very clear and important guide for the formulation of our mechanism of vacuum energy sequestration, to which we will turn in the next section.

Consider a local and locally Poincar\'e-invariant $4D$ QFT  describing finitely many degrees of freedom below a certain UV cutoff $M_{UV}$. Imagine, for starters, that it couples minimally to gravity described by the $4D$ metric $g_{\mu\nu}$. Next look for a Poincar\'e-invariant vacuum; if it exists, the theory admits a vacuum with a zero cosmological constant. Clearly, the aim is to evaluate under which circumstances this can happen. Now, in the Poincar\'e state, all the fields, regardless of their spin are annihilated by the translation generators. Choosing them as the coordinate basis, this implies that the fields must be $\Phi_m =const$
and $g_{\mu\nu}= \eta_{\mu\nu}$ modulo a residual rigid GL(4) symmetry, inherited from diffeomorphisms.
Hence the field equations for matter and gravity, respectively, reduce to 
\be
\frac{\del    {\cal L}}{\del \Phi_m} = 0\, , ~~~~~~~~~~~~ \frac{\del    {\cal L}}{\del g_{\mu\nu}} =0 \, .
\label{eqw2}
\ee
The residual rigid GL(4) symmetry of the system is realized linearly. The metric and matter fields and
the Lagrangian density transform as $g_{\mu\nu} \rightarrow J_{\mu}{}^\alpha J_{\nu}{}^\beta g_{\alpha \beta}$, $\Phi_m \rightarrow {\cal J}(J) \Phi_m$ and ${\cal L} \rightarrow \det(J) {\cal L}$, respectively, where ${\cal J}(J)$ is some appropriate representation of GL(4).
Now, since the field equations (\ref{eqw2}) are linearly independent, and the Poincare-invariant ground state is unique, by virtue of the matter field equations in (\ref{eqw2}) and the residual GL(4) symmetry, the Lagrangian in this state is
\be
{\cal L} = \sqrt{\eta} \Lambda_0(\Phi_m) \, ,
\label{ccterm}
\ee
where $\Phi_m$ are field configurations which extremize $\Lambda_0$. Here, clearly, one can think of 
$\Lambda_0$ and $\Phi_m$ as the renormalized variables, computed to some fixed order in the loop expansion, and using some type of a covariant regulator and subtraction scheme to cancel the divergent contribution and obtain a finite value. 
The bottomline is that the final equation of (\ref{eqw2})
then implies $\Lambda_0 \eta_{\mu\nu} = 0$, and so consistency requires setting $\Lambda_0 = 0$ -- by hand.
And resetting it to zero -- also by hand -- if further loop corrections are included.
Technically, the problem is that the last Eq. of (\ref{eqw2}) is completely independent of the other equations.

\begin{figure*}[h]
\centering
\includegraphics[height=8.5cm,width=12cm]{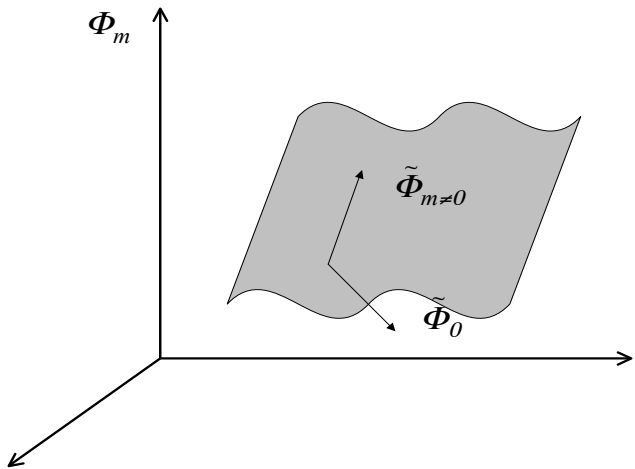}
\caption{Field redefinition $\Phi_m \rightarrow \tilde \Phi_m$.}
\label{frobenius}
\end{figure*}

What if that were not the case? Clearly, an adjustment mechanism, if it exists, would set the value of $\Lambda_0$ automatically to zero (or sufficiently close to it) once field variables attain their extrema.
This could be enforced by requiring that the trace of the last Eq. of (\ref{eqw2}) is replaced by an equation of the form
\be
2 g_{\mu\nu} \frac{\del {\cal L}}{\del g_{\mu\nu}} =
\sum_m f_m(\Phi_{n})
\frac{\del {\cal L}}{\del \Phi_m} \, .
\label{cccond}
\ee
If a theory exists such that one of the resulting gravitational equations yields (\ref{cccond}), it would open the road to constructing a successful adjustment mechanism within the domain of effective QFT coupled to gravity.
Furthermore, to ensure the absence of fine tunings, one can require that  $f_m(\Phi_{ n})$ are a set of smooth functions in field space, which only depend on the fields $\Phi_{ m}$ in order to maintain 
Poincar\'e invariance. A closer look \cite{wein} reveals that the first order (functional)  
partial differential equation (\ref{cccond}) is in fact a requirement that the total action describing the theory 
is invariant under a transformation generated by
\be
\delta g_{\mu\nu} = 2 \epsilon g_{\mu\nu} \, ,
~~~~~~~~~~~~~~ \delta \Phi_m = - \epsilon f_m(\Phi_{n}) \, .
\label{symm4d}
\ee
This is clearly a scaling symmetry in disguise. Now, since (1) the number of field 
theory degrees of freedom is finite, and (2) the functions $f_m(\Phi_n)$ are smooth,
one can perform a field redefinition $\Phi_m \rightarrow \tilde
\Phi_m = \tilde \Phi_m(\Phi_n)$, so that in terms of the new field theory degrees of freedom the transformations 
(\ref{symm4d}) simplify to
\be
\delta g_{\mu\nu} = 2 \epsilon g_{\mu\nu} \, ,
~~~~~~~~~~ \delta \tilde \Phi_0 = - \epsilon  \, ,
~~~~~~~~~~ \delta \tilde \Phi_{m \neq 0} = 0 \, .
\label{symm4dnew}
\ee
That such a transformation exists is guaranteed by a theorem of differential geometry on embeddings of hypersurfaces \cite{goldbish} and Poincar\'e symmetry. The field space generator of transformations (\ref{symm4d}) is a smooth finite-dimensional vector field $X = \sum_m f_m(\Phi_{n}) \frac{\del}{\del \Phi_m}$, so that the transformation (\ref{symm4d}) represents a motion along the flow lines generated by $X$. Then one defines the new field space coordinates by picking the parameter measuring the flow along $X$
(which being a single degree of freedom must be a scalar by unbroken local Poincare invariance)
and the coordinates orthogonal to it.  They are the $n-1$ integration constants $\tilde \Phi_{m\ne1}$ of the solution of the system of
differential equations $\frac{\delta \Phi_m}{\delta \tilde \Phi_0} = f_m(\Phi_{n})$ which remain after using one to pick the origin of $\tilde \Phi_0$.
See Figure \ref{frobenius}.

In the new coordinates in field space, the field theory equations, the Poincare-invariance of ground state and the residual GL(4) symmetry imply that the Lagrangian in this state is
\be
{\cal L} = \sqrt{\eta}  \Lambda_0(\tilde \Phi_{m \neq 0} ) e^{-4 \tilde \Phi_0} \, .
\label{runsoln}
\ee
Again, this is the regulated and renormalized vacuum energy coming from a calculation involving contributions from all diagrams in the loop expansion involving some fixed, but arbitrarily chosen, finite number of loops. Now, the modified gravity equation (\ref{cccond}) yields $\Lambda_0(\tilde \Phi_{m \neq 0} ) e^{-4 \tilde \Phi_0}=0$. This could be solved while avoiding fine tuning $\Lambda_0 = 0$ if one allows $\tilde \Phi_0$ to run off to infinity. However: because Eq. (\ref{runsoln}) is the renormalized vacuum energy at an arbitrary finite order in the loop expansion, and since the mechanism suppressing it must operate independently  of the order of the loop expansion to guarantee radiative stability, the form of 
(\ref{runsoln}) must be preserved order-by-order in perturbation theory. But this means that in order to satisfy this, {\it all} the scales in the theory must depend on the power of $\exp(-\tilde \Phi_0)$ given by their engineering dimension. This includes the scales of the regulator and the subtraction point, and is necessary because the corrections to (\ref{runsoln}) come as the powers of these scales and logarithms of their ratios. Only then will all the corrections, including the 
log-dependent terms, will scale the same 
way as in (\ref{runsoln}). 
So (\ref{runsoln}) will accomplish the task if all the matter fields and the regulator couple to the `rescaled' metric $\hat g_{\mu\nu} = \exp(-2\tilde \Phi_0) g_{\mu\nu}$. However, after canonically normalizing the fields in the matter sector on the background given by solving the vacuum equations, one finds that all dimensional parameters in QFT must scale as $m^d \propto e^{-d\tilde \Phi_0}$. This means that in the limit
$\tilde \Phi_0 \rightarrow \infty$ not only does the cosmological constant vanish, but so do all the other scales in the theory! In other words, this restores conformal symmetry in the field theory sector. As noted this is not our world. Hence the problem.

\section{Our proposal} \label{prop}

In the attempt to evade the Weinberg's no-go theorem, in \cite{kalpad1} we proposed a very minimal modification of General Relativity with minimally coupled matter. The idea was to promote the classical cosmological constant $\Lambda$ to that of a global dynamical variable, and introduce a second global variable $\lambda$ corresponding to scales in the matter sector. The variation with respect to these new variables however is used to impose a global constraint on the dynamics of the theory, similar in spirit to the constraint imposed in the classic isoperimetric problem of variational calculus 
\cite{elsgolts}. To do this, we supplemented the local action  with an additive function $\sigma(\Lambda, \lambda)$ which is not integrated over the space time. Then the variations with respect to $\Lambda,\lambda$ select the values of these parameters. In particular, this procedure sets the boundary condition for the variable $\lambda$ such that at any order of the loop expansion is takes precisely the right value to completely absorb away the whole of vacuum energy contribution from the matter sector at that loop order. Our approach is a simplified hybrid of thinking about GR as unimodular gravity, with a variable $\Lambda$ specified by arbitrary cosmological initial conditions, and the proposal of Linde \cite{linde2}, further considered by Tseytlin \cite{tsey}, of using modified variational procedures to fix values of global variables such as $\Lambda$. Our variational prescription however differs from those previously considered in that it is more minimal, and that it uses a global scaling symmetry
as an organizing principle for accounting for all quantum vacuum energy contributions.

The idea is to start with the action
\be
S= \int d^4 x \sqrt{g} \left[ \frac{M^2_{Pl}}{2} R  - \Lambda - {\lambda^4} {\cal L}(\lambda^{-2} g^{\mu\nu} , \Phi) \right] +\sigma\left(\frac{ \Lambda}{\lambda^4 \mu^4}\right) 
 \, ,
 \label{action}
\ee
where the matter sector described by ${\cal L}$ is minimally coupled to the metric $\tilde g_{\mu\nu}=\lambda^2 g_{\mu\nu}$. We imagine that the Standard Model is included in it. For simplicity, we consider the matter to only belong to this sector, henceforth referred to as the `protected sector'. One could also include other matter sectors to the theory, that could couple to a different combination of $\lambda$ and $g_{\mu\nu}$. However, vacuum energy contributions from such sectors would not cancel automatically. In what follows we will focus on the matter dynamics only from the protected sector, for simplicity's sake. The function $\sigma(z)$ is a (odd) differentiable function which imposes the global constraints.  The parameter $\mu$ is a mass scale introduced on dimensional grounds. The precise form of $\sigma$ is determined in order to fix the particle masses in QFT in accordance with a specific phenomenological model of physics beyond the Standard Model, as we will discuss in Sec. \ref{sec:masses}. Eg., an asymptotically exponential form of $\sigma$ allows us to choose $\mu$ to be up near the Planck scale etc.

The global variable $\lambda$ sets the hierarchy between the matter scales and the Planck scale, since 
\be
\frac{m_{phys}}{M_{Pl}} =  \lambda \frac{m}{M_{Pl}} \, ,
\label{scaling}
\ee
 where $m_{phys}$ is a physical mass scale of a canonically normalized matter theory, and $m$ is the bare mass in the Lagrangian. As an illustration, consider a scalar field with bare mass $m$, 
 \ba
\sqrt{\tilde g}{\cal L}_\phi &=& \frac{1}{2}\sqrt{\tilde g }\left[\tilde g^{\mu\nu} \del_\mu \phi \del_\nu \phi+m^2 \phi^2\right]  \nonumber\\
&=&    \frac{1}{2}  \sqrt{g} \lambda^4 \left[\lambda^{-2} g^{\mu\nu} \del_\mu \phi \del_\nu \phi+m^2 \phi^2 \right] \nonumber \\
&=&    \frac{1}{2}  \sqrt{g} \left[ g^{\mu\nu} \del_\mu \varphi \del_\nu \varphi+m_{phys}^2 \varphi^2 \right] \, ,
\ea
where $\varphi=\lambda \phi$ is the canonical scalar and  the physical mass is  $m_{phys}=\lambda m$.

It is absolutely essential for our cancellation mechanism to enforce the UV regulator of this sector to also couple to exactly the same metric as the fields from ${\cal L}$.
This is necessary in order to ensure the correct operational form of the vacuum energy sequestration from 
${\cal L}$, and makes the effective UV cutoff $M_{UV}$ and the subtraction scale ${\cal M}$ scale with $\lambda$ in exactly the same way as mass scales from ${\cal L}$, given by Eq. (\ref{scaling}).
This can be accomplished, for example, by regulating the theory with a system of Pauli-Villars regulators, which couple to the metric $\tilde g_{\mu\nu}$.

With this in place, the form of the (\ref{action}) {\it guarantees} that {\it all} vacuum energy contributions coming from
the protected Lagrangian $\sqrt{g} \lambda^4 {\cal L}(\lambda^{-2} g^{\mu\nu} , \Phi)$ must depend on $\lambda$ only through an overall scaling by $\lambda^4$,
even after the logarithmic corrections are included. This follows since the regulator of the QFT introduces contributions where the scales depend on $\lambda$ in the same way as those from the physical degrees of freedom from ${\cal L}$. That ensures the cancellation of $\lambda$ in loop logarithms, and so only the powers of $\lambda$ remain. This fact follows from diffeomorphism invariance of the theory, which since $\lambda$ is a global variable, is unbroken both for the metrics $g_{\mu\nu}$ and $\tilde g_{\mu\nu}$. Diffeomorphism 
invariance ensures that the full effective Lagrangian computed from 
$\sqrt{g} \lambda^4 {\cal L}(\lambda^{-2} g^{\mu\nu} , \Phi)=\sqrt{\tilde g} {\cal L}(\tilde g^{\mu\nu} , \Phi)$, 
including all quantum corrections -- and the divergent terms, as well -- still couples to the exact same $\tilde g_{\mu\nu}$ \cite{selft}.  Of course, this is true only when we restrict our attention to the vacuum energy contributions from loop diagrams involving only protected sector degrees of freedom in the internal lines which are integrated over. This is, however, all of the vacuum energy in the decoupling limit of gravity, which we calculate using standard flat space field theory techniques in locally freely falling frames. 
As we stressed previously, we re-couple these terms to gravity by the minimal coupling procedure, taking gravity  as a purely (semi) classical field which merely serves the purpose of {\it detecting} vacuum energy. At present, this is the sharply formulated part of the cosmological constant problem, and as we discussed in the introduction we choose to focus on it alone.

The field equations that follow from varying the action (\ref{action}) with respect to global auxiliary fields $\Lambda, \lambda$ are
\be
\frac{\sigma'}{\lambda^4\mu^4} = \int d^4x \sqrt{g} \, , ~~~~~~~~~~~~~~~~ 4\Lambda \frac{ \sigma' }{\lambda^4\mu^4} 
= \int d^4x \sqrt{g} \,\lambda^4 \, \tilde T^{\mu}{}_\mu \, ,
\label{varsl}
\ee
where $\tilde T_{\mu\nu}=-\frac{2}{\sqrt{\tilde g}} \frac{\delta S_m}{\delta \tilde g^{\mu\nu}}$ is the energy-momentum tensor defined in the `Jordan frame'. To go to the `physical' frame, in which matter sector is canonically normalized, note that
 $$T^\mu{}_\nu=g^{\mu\alpha}\left[-\frac{2}{\sqrt{ g}} \frac{\delta S_m}{\delta  g^{\alpha\nu}} \right] =\lambda^2 \tilde g^{\mu\alpha} \left[-\frac{2\lambda^4 }{ \sqrt{\tilde  g}} \frac{\delta S_m}{\lambda^{2} \delta  \tilde g^{\alpha\nu}} \right] =\lambda^4 \tilde T^\mu{}_\nu \, ,$$
where $\sigma' = \frac{d\sigma(z)}{dz}$. As long as it is nonzero\footnote{And non-degenerate: it can't be the pure logarithm. In that case Eqs. (\ref{varsl}) turn into two independent constraints,
$\frac{1}{\Lambda} = \int d^4x \sqrt{g},~4
= \int d^4x \sqrt{g} \, T^{\mu}{}_\mu $, the latter placing an artificial constraint on the matter sector.}, eliminating it from the two Eqs. (\ref{varsl}) one finds
 $$\Lambda = \frac14 \langle T^\mu{}_\mu \rangle \, ,$$
where we defined the  $4$-volume average of a quantity by $\langle Q\rangle ={\int d^4 x \sqrt{g} \, Q}/{\int d^4 x \sqrt{g}}$. Note that $\Lambda$ is the bare cosmological constant, which is now however completely fixed by this condition. 
One has to define these averages meaningfully, since after regulating the divergences they can still be indeterminate ratios when the space-time volume is infinite \cite{linde}. We will address this in short order, since this issue has very important physical implications and ties into how our proposal evades Weinberg's no-go theorem.

The variation of (\ref{action}) with respect to $g_{\mu\nu}$ yields 
\be \label{ein}
M_{Pl}^2 G^\mu{}_\nu = -\Lambda \delta^\mu{}_\nu + \lambda^4 \tilde T^\mu{}_\nu \, ,
\ee
where $G^{\mu}{}_\nu$ is the standard Einstein tensor. After eliminating $\Lambda$ and canonically normalizing the matter sector, this becomes
\be
M_{Pl}^2 G^\mu{}_\nu = T^\mu{}_\nu-\frac{1}{4} \delta^\mu{}_\nu \langle   T^\alpha{}_\alpha \rangle \, ,
\label{eeqs}
\ee
This equation is  one of the two key ingredients of our proposal. Note that this is  the full system of {\it ten} field equations, with  the trace equation {\it included}.  It differs from 
unimodular gravity \cite{finkel,buch,henteitel,unruh,ng,kuchar,ellis,barrow,ian}, where although the restricted variation removes the trace equation that involves the vacuum energy, this equation comes back along with an arbitrary integration constant, after using the Bianchi identity. Here
there are {\it no} hidden equations nor integration constants,  and all the sources are automatically accounted for in (\ref{eeqs}).  Crucially, however,  
$$\frac14 \langle  T^\alpha{}_\alpha \rangle$$ 
is subtracted from the right-hand side of (\ref{eeqs}). This means that the hard cosmological constant, be it a classical contribution to ${\cal L}$ in (\ref{action}), or a quantum vacuum correction calculated to any order in the loop expansion, divergent (but regulated!) or finite, {\it never} contributes to the field equations (\ref{eeqs}). To see this explicitly, we  take the effective matter Lagrangian, ${\cal L}_\textrm{eff}$ at any given order in loops, and split it into the 
renormalized quantum
vacuum energy contributions (classical and quantum) $\tilde V_{vac}=\langle 0| {\cal L}_\textrm{eff} (\tilde g^{\mu\nu}, \Phi )|0\rangle$, and local excitations $\Delta {\cal L}_\textrm{eff}$, 
\be
\lambda^4 \sqrt{g} {\cal L}_\textrm{eff}(\lambda^{-2} g^{\mu\nu} , \Phi)=\lambda^4 \sqrt{g}\left[ \tilde  V_{vac} + \Delta {\cal L}_\textrm{eff}(\lambda^{-2} g^{\mu\nu} , \Phi)\right] \, .
\ee
It follows that $T^{\mu}{}_\nu = V_{vac} \delta^\mu{}_\nu + \tau^\mu{}_\nu$, where $V_{vac}=\lambda^4\tilde V_{vac} $ is the total regularized vacuum energy and $\tau_{\mu\nu}= \frac{2}{\sqrt{g}} \frac{\delta}{\delta g^{\mu\nu}} \int d^4x \sqrt{g} \lambda^4 \Delta {\cal L}_\textrm{eff}(\lambda^{-2} g^{\mu\nu}, \Phi)$ describes the physical excitations. By our definition of the historic 4-volume  average, 
$\langle V_{vac} \rangle \equiv  V_{vac}$ and so  the field equations (\ref{eeqs}) become
\be 
M_{Pl}^2 G^\mu{}_\nu=  \tau^\mu{}_\nu-\frac{1}{4} \delta^\mu{}_\nu \langle  \tau^\alpha{}_\alpha \rangle \, ,
\label{eeqs1}
\ee
The regularized vacuum energy $V_{vac}$  has completely dropped out from the source in (\ref{eeqs}). There is a residual effective cosmological constant coming from the historic average of the trace of matter excitations: 
\be
\Lambda_\text{eff} =\frac14 \langle \tau^\alpha{}_\alpha \rangle \, .
\label{residual}
\ee
We emphasize that this residual cosmological constant has absolutely nothing to do with the vacuum energy contributions from the matter sector, including the Standard Model contributions. 
Instead, after the cancellation of the vacuum energy contributions enforced by the global variables $\Lambda, \lambda$, the residual value of $\langle \tau^\alpha{}_\alpha \rangle$ is picked as a `boundary condition', 
resulting from the measurement of the finite part of the cosmological constant after it was renormalized. As we noted above, this measurement requires the whole history of the universe for this one variable, in effect setting the boundary condition for it at future infinity. Obviously, it is crucial that the numerical value of $\Lambda_\textrm{eff}$ is automatically small - one might hope for it, since it is the contribution coming predominantly from the IR modes in the cosmological evolution, but a quantitative confirmation is necessary since otherwise the proposal would have failed. It turns out that indeed
$\Lambda_\textrm{eff}$ is automatically small enough in large old universes, as we will show in the next section.
This follows from the fact that our universe is large and old, which in the very least is a consequence of extremely weak anthropic considerations. 
In actual fact, a large universe like ours can be formed  by at least 60 efolds of inflation, which we will show is consistent with the vacuum sequestration proposal. The smallness of $\Lambda_\textrm{eff}$ is   completely safe from radiative instabilities, and the required dynamics is essentially insensitive of the order of perturbation theory. The apparent acausality in the determination of $\Lambda_\textrm{eff}$ is therefore of no consequence; a better terminology is {\it aposteriority}, that follows from the nature of the measuring process needed to set the numerical value of the renormalized cosmological constant. This has no impact on local physics and so cannot lead to any pathologies normally associated with any local acausality. One might only worry if there do not appear some unexpected restrictions on a possible range of numerical values which the renormalized cosmological constant might have, that follow from the global constraints which we introduced. We will turn to this later, when we address how the solutions of the theory are constructed.

The second key ingredient of our mechanism is that the field theory spectrum has a nonzero gap, which can be arbitrarily large compared to $|\langle \tau^\alpha{}_\alpha \rangle|^{1/4}$. Otherwise, the mechanism would fail to provide a way around Weinberg's no-go theorem - reducing yet again to a framework with at least scaling symmetry. 
This means that on the solutions the parameter $\lambda$ must be nonzero,  since  
$\lambda \propto m_{phys}/M_{Pl}$. But by virtue of the first of Eqs. (\ref{varsl}), since $\sigma(z)$ is a differentiable function, if $\lambda$ is nonzero, $\int d^4x \sqrt{g}$ must be {\it finite}. Fortunately this can be accomplished in a universe with spatially compact sections, which is also temporally finite: it starts with a Bang and ends with a Crunch. In other words, the spacelike singularities regulate the worldvolume of the universe without destroying diffeomorphism invariance and local Poincar\'e symmetry. Therefore, in our framework the universes which support non-scale invariant particle physics must be spatiotemporally finite. Infinite universes are solutions too, however their phenomenology is not a good approximation to our world, since all scales in the protected sector vanish. Note, that the quantity which controls the value of $\lambda$, and therefore 
$m_{phys}/M_{Pl}$, is the worldvolume of the universe, and not $\langle \tau^\alpha{}_\alpha \rangle$. This is essential for the phenomenological viability of our proposal, since it separates the scales of the observed cosmological constant and masses of particle physics.

We will address these points in more detail  in section  \ref{sec:hist}, focussing on quantitative statements,  demonstrating, in particular,  that the residual cosmological constant is naturally small, never exceeding the critical density of the universe today. This is of course the final {\it touch\'e} of our model, guaranteeing that the cancellations of the vacuum energy did not -- in turn -- necessitate large classical values to appear. Before we address these, and other phenomenologically important issues, let us look in more depth at just how the vacuum energy contributions get cancelled.

A holy grail of the past attempts to protect the cosmological constant from radiative corrections, which is partially realized with supersymmetry and conformal symmetry, was to find a symmetry which will insulate the finite term left after renormalization from higher order loop corrections. As it turns out, a system of two symmetries is the reason why our cancellation works. 
Our action (\ref{action}) has approximate scale invariance,  broken only by the Einstein-Hilbert term,  
\be
\lambda \rightarrow \Omega \lambda \, ,\qquad 
g_{\mu\nu} \rightarrow \Omega^{-2} g_{\mu\nu} \, , \qquad\Lambda \rightarrow \Omega^4 \Lambda \, ,
\label{scale}
\ee
such that the action changes by 
\be
\delta S =\frac{M^2_{Pl}}{2} \Omega^{-2} \int d^4 x \sqrt{g} R= \frac{M^2_{Pl}}{2} \Omega^{-2} \langle R \rangle \int d^4 x \sqrt{g} \, .
\label{scalchange}
\ee
The other symmetry appears as an approximate shift symmetry
\be
{\cal L} \to {\cal L}+\epsilon m^4 \, , \qquad \Lambda \to \Lambda - \epsilon  \lambda^4 m^4  \, ,
\label{shiftsymmetry}
\ee
under which the action changes by
\be
\delta S = \sigma\left(\frac{\Lambda}{\lambda^4 \mu^4} - \epsilon \frac{m^4}{\mu^4}\right) - \sigma\left(\frac{\Lambda}{\lambda^4 \mu^4}\right)
\simeq - \epsilon \sigma' \frac{m^4}{\mu^4} \, .
\label{shiftchange}
\ee
The scaling symmetry ensures that the vacuum energy at an arbitrary order in the loop expansion couples to gravitational sector exactly the same way as the classical piece. The `shift symmetry' of the bulk action then cancels the matter vacuum energy and its quantum corrections. The scaling symmetry breaking by the gravitational sector is mediated to the matter only by the cosmological evolution,
through the scale dependence on $\int d^4 x \sqrt{g}$, and so is weak.
This is why the residual cosmological constant is {\it small}: substituting the first of Eqs. (\ref{varsl}) and using $\lambda = m_{phys}/m$, we see that $\delta S \simeq - \epsilon m^4 \lambda^4 \int d^4 x \sqrt{g} = - \epsilon  \left(\frac{m_{phys}}{M_{Pl}}\right)^4 \left[ M_{Pl}^4 \int d^4 x \sqrt{g} \right] $. Holding the volume fixed, this is small when $m_{phys}/M_{Pl} \ll 1$, vanishing in the conformal limit\footnote{Defined by fixing $\int d^4 x \sqrt{g}$ and taking $\mu \rightarrow \infty$ in the first of 
Eqs. (\ref{varsl}).} $\lambda \propto m_{phys} \rightarrow 0$. This restoration of symmetry renders
a small residual curvature technically natural.

Let us see how this occurs at the level of the field equations (\ref{varsl}), (\ref{eeqs}). First off, note that the historic average of the trace of the right hand side of (\ref{eeqs}) is zero, $\langle R \rangle =0$. Hence, by (\ref{scalchange}), the action
(\ref{action}) is in fact invariant under scaling (\ref{scale}) on shell. Next, the variation of the matter Lagrangian under the shift ${\cal L} \to {\cal L}+\epsilon m^4$ is equivalent to letting $\tilde T^\mu{}_{\nu} \to  \tilde T^\mu{}_{\nu} -\epsilon m^4 \delta^\mu{}_{\nu}$. Then if $g_{\mu\nu}$, $\Lambda$ and $\lambda$ solve Eqs. (\ref{varsl}) and (\ref{ein}) for a source $\tilde T^\mu{}_{\nu} $ , by manipulating the constraint equations (\ref{varsl}) one finds that 
\be
\hat g_{\mu\nu}=g_{\mu\nu}, \qquad \hat \Lambda=\Lambda \frac{\hat z \sigma'(\hat z)}{z\sigma'(z)}, \qquad \hat \lambda^4=\lambda^4 \frac{\sigma'(\hat z)}{\sigma'( z)} \, ,
\ee
are solutions with a source $\tilde T^\mu{}_{\nu} -\epsilon m^4 \delta^\mu{}_{\nu}$, where 
$z=\Lambda/\lambda^4 \mu^4$, and $\hat z=z\left(1-4\epsilon m^4/\langle \tilde T \rangle\right)$.
Crucially, $g_{\mu\nu}$ remains unchanged, which means that a shift of the vacuum energy -- by adding higher order corrections in the loop expansion -- is absorbed by an automatic readjustment of 
the global variables, and so has no impact whatsoever on the geometry.

Finally, since the role of the two global  auxiliary fields, $\lambda$ and $\Lambda$ is to enforce global constraints, they only appear algebraically in the theory and can be integrated out explicitly. The resulting formulation of the theory provides further insight into the protection mechanism. So using the variable $z$ defined above, the constraints (\ref{varsl}) are 
\be
\frac{\sigma'(z)}{\lambda^4 \mu^4}=\int d^4 x \sqrt{g} \, , \qquad z=\frac{\langle \tilde T^\alpha{}_\alpha\rangle}{4\mu^4} \, .
\ee
Solving for $\lambda, \Lambda$ yields
\be \label{lamLam}
\lambda=\left[\frac{\sigma'\left(\langle \tilde T^\alpha{}_\alpha\rangle /4 \mu^4  \right)}{\mu^4 \int d^4x \sqrt{g} }\right]^{1/4} \, , \qquad \Lambda=\frac{\sigma'\left(\langle \tilde T^\alpha{}_\alpha\rangle /4 \mu^4  \right) \langle \tilde T^\alpha{}_\alpha\rangle /4 \mu^4 }{\int d^4 x \sqrt{g} } \, .
\ee
Substituting these back into the action, we obtain
\be
S= \int d^4 x \sqrt{g} \left[ \frac{M^2_{Pl}}{2} R   - {\lambda^4} {\cal L}(\lambda^{-2} g^{\mu\nu} , \Phi) \right] +F\left(\langle \tilde T^\alpha{}_\alpha\rangle /4 \mu^4  \right)
 \, , ~~
\label{redaction}
\ee
where $\lambda$ is given by equation (\ref{lamLam}), and $\tilde T^\alpha{}_\alpha=\tilde g^{\alpha \beta} \left(\frac{2}{\sqrt{\tilde g}} \frac{\delta}{\delta \tilde g^{\alpha \beta}} \int \sqrt{\tilde g} {\cal L}(\tilde g^{\mu\nu} , \Phi)\right)$. An important point here is that 
the function $F(z)=\sigma(z)-z\sigma'(z)$ is the {\it Legendre transform} of $\sigma$. 
This explicitly shows that the independent variable $z$ (or $\Lambda$) has been traded for the {\it new} independent variable $\sigma'(z)$ -- which, by the first of Eqs. (\ref{lamLam}), is $\lambda^4 \mu^4 \int d^4x \sqrt{g} \,\,$. Thus, the independent variable of the theory is really {\it not} the cosmological counterterm $\Lambda$,  (like in GR or in unimodular extension of GR)
but the worldvolume of the universe, $\int d^4x \sqrt{g}$. So this in fact reveals how our mechanism operates. The cosmological system  responds instead to changes in the space-time volume, rather than changes in $\Lambda$. Furthermore, in GR, when higher order corrections to the vacuum energy are included, $\Lambda$ stays fixed forcing the space-time volume $\int d^4 x \sqrt{g}$ to absorb the corrections (inflating a lot due to a large vacuum energy). For our scenario it is exactly the opposite: it is the space-time volume that remains fixed, forcing $\Lambda$ to adjust. As a result, the cosmological system is stable against radiative corrections to the vacuum energy.

Indeed, let us consider the gradient expansion of the effective matter Lagrangian, including any number of loop corrections. To the lowest order, we retain only the full effective potential of the theory, and truncate it to the zero momentum limit, which represents the vacuum energy $\tilde V_{vac} $.
We find that ${\cal L}_\textrm{eff}=\tilde V_{vac}, ~\langle \tilde T^\alpha{}_\alpha \rangle=-4\tilde V_{vac}$, and the action is 
\ba
S &=& \int d^4 x \sqrt{g} \left[ \frac{M^2_{Pl}}{2} R   -  \left(\frac{\sigma'\left(-\tilde V_{vac} /\mu^4  \right)}{\mu^4 \int \sqrt{g} }\right) \tilde V_{vac}  \right] +F\left(-\tilde V_{vac}/\mu^4  \right)
\nonumber  \\
&=&   \int d^4 x \sqrt{g} \left[ \frac{M^2_{Pl}}{2} R    \right] +\sigma\left(-\tilde V_{vac}/\mu^4  \right) \, .
\ea
We see how the choice of coupling of $g_{\mu\nu}$ to the Standard Model given by equation (\ref{redaction}) guarantees that all $g_{\mu\nu}$ dependence is cancelled in the vacuum energy contribution to the action. This is how the Standard Model vacuum energy is sequestered. Diffeomorphism invariance guarantees that the full effective Lagrangian computed from $\sqrt{\tilde g} {\cal L}(\tilde g^{\mu\nu} , \Phi)$, including all quantum corrections, still couples to the same $\tilde g_{\mu\nu}=\left[\frac{\sigma'\left(\langle \tilde T^\alpha{}_\alpha\rangle /4 \mu^4  \right)}{\mu^4 \int \sqrt{g} }\right]^{1/2}g_{\mu\nu}$. The loop corrections are accommodated by (small!) changes of the scaling factor 
$\left[\frac{\sigma'\left(\langle \tilde T^\alpha{}_\alpha\rangle /4 \mu^4  \right)}{\mu^4 \int \sqrt{g} }\right]^{1/2}$, while the metric $g_{\mu\nu}$ remains completely unaffected. This works to any order in the loop expansion, as is the core element of the adjustment mechanism.

So to summarize, we see that the key point of our mechanism is a dramatically altered role of  $\int d^4x \sqrt{g}$, which provides the way around Weinberg's {\it no-go} theorem \cite{wein}. Instead of $\Lambda$ in GR or in its unimodular formulation, now $\int d^4x \sqrt{g}$ is the independent variable. Taking all the physical scales in the protected matter sector to depend on it, automatically removes all the vacuum energy contributions to drop out. For example, if we were to take the linear function $\sigma(z)=z$ in (\ref{action}) 
and declare ${\cal L}$ to be literally the Standard Model, integrating $\Lambda$ and $\lambda$ out and rewriting (\ref{action}) as just Einstein-Hilbert action coupled 
to the Standard Model, the only modification is that the Higgs vev $v$ is replaced by $v/(\mu^4 \int d^4 x \sqrt{g})^{1/4}$! Further, in a collapsing spacetime, since $\int d^4x \sqrt{g}$ is finite, the protected sector QFT has a nonzero mass gap, by virtue of (\ref{varsl}), while the residual cosmological constant $\langle \tau^\alpha{}_\alpha \rangle/4 \ne 0$, but it is completely independent of all the vacuum energy corrections, and as we will show below is automatically small in a large old universe. The particle sector scales are affected by the historic value of the worldvolume\footnote{In the case where $\sigma$ is a linear function, they would be too sensitive to the initial conditions in the early universe. This is why the forms of $\sigma$ which are asymptotically exponential are preferred, since they reduce the sensitivity to merely logarithmic corrections.}. However, this dependence 
of field theory scales on $\int d^4x \sqrt{g}$ is completely invisible to any nongravitational local experiment, by diffeomorphism invariance and local Poincar\'e symmetry. Locally the theory looks just like standard GR, in the (semi) classical limit - but without a large cosmological constant, and without its radiative instability, at least in the limit of (semi) classical gravity. 

\section{Historic integrals and quantitative considerations} \label{sec:hist}

We have already stated that in universe which is compact, starting in a Bang and ending in a Crunch, the worldvolume $\int d^4 x \sqrt{g}$ is finite. Also the residual cosmological constant given by the historic average of the trace of stress-energy tensor $-\langle \tau^\alpha{}_\alpha \rangle/4$
is automatically small in big and large universes, as long as $\tau^\mu{}_\nu$ satisfies the  dominant and null energy conditions (DEC and NEC, respectively). In this section we will give a proof of this statement at the classical level. 
Semiclassically, it has been shown that the integrals $\langle \tau^\alpha{}_\alpha \rangle$ are finite if NEC is valid in \cite{wald}. Then for bounded $\tau^\alpha{}_\alpha$ our argument automatically extends to the semiclassical case, too. 
Further, we will consider the techniques for finding solutions
of field equations in our setup. From a practical point of view, solving the equation (\ref{eeqs1}) requires a bootstrap method: allow $\langle  \tau^\alpha{}_\alpha \rangle = C$ to be arbitrary to start with, find the family of solutions parameterized by this integration constant, and finally substitute this family of solutions back into $\langle  \tau^\alpha{}_\alpha \rangle$ in order to show that  a subset of them is compatible with the initial choice. This is akin to the gap equation in superconductivity. Here we will focus on the proof of existence of solutions to this procedure in the case of background FRW cosmologies. In the forthcoming work \cite{KP3} we will show that consistent solutions of this type including the current epoch of (transient) acceleration exist. Specific dynamics will resort to the earliest quintessence with linear potential dating back to, at least, 1987 \cite{andrei87}, as the really relevant model of transient cosmic acceleration. We will show in \cite{KP3} that it has a natural embedding in our proposal. Other scenarios and aspects of transient acceleration have also been considered (see eg \cite{others}).
 
\subsection{Historic integrals and the cosmological background} \label{sec:histint}

The integrals which appear in the definition of our historic averages are
\ba
\int d^4 x \sqrt{g}&\supset &\text{Vol}_3 \int_{t_{bang}}^{t_{crunch} }dt a^3 \, , \label{ints1} \\
\int d^4 x \sqrt{g}\tau^\alpha{}_\alpha &\supset &\text{Vol}_3 \int_{t_{bang}}^{t_{crunch}} dt a^3 (-\rho+3p) \, ,
\label{ints}
\ea
where the cosmology takes place over a finite proper time interval $t_{bang}<t<t_{crunch}$, with a scale factor $a$, and finite spatial comoving volume $\text{Vol}_3$.  We will approximate the integrals for the most part by the FRW geometry with fluids, whose energy density and pressure are 
$\rho$ and $p$ respectively. We will eventually impose DEC and NEC, which combined together require $|p/\rho| \le 1$.  The integrals are regulated - ie., finite - because we assume that spatial sections are compact, and that the universe starts at a Bang and ends in Crunch. This guarantees that (\ref{ints1}) is finite, and that the QFT spectrum has a nonzero mass gap. Further, we will see that validity of NEC then also guarantees that (\ref{ints}) is also finite, and bounded by the contributions from near the turning point, being estimated by the product of the minimal energy density during cosmological evolution and the age of the universe. At the technical level, we will take the evolution to be symmetric in time for simplicity. This does not impair the generality of our analysis because the integrals are dominated by the contributions near the turning point. Approximating the spatial volumes by homogeneous 3-geometries, we will first consider time integrals. We will briefly return to contributions from inhomogeneities later, when we consider the effects from black holes.

For starters, note that the DEC and NEC together -- $|p/\rho | \le 1$ -- put a very strong bound on the contributions to the integral (\ref{ints}) from the Bang and Crunch singularities. Namely while at the singularities the energy density and pressure diverge, the spacelike volume which they occupy shrinks, and $|p/\rho | \le 1$ ensures that the rate of shrinking is faster than the divergence of $\tau^\alpha{}_\alpha$. Indeed, near the singularities $\rho$ scales as $\rho \sim 1/(t-t_{end})^2$, by virtue of the Friedman equation, where $t_{end}$ is either of the singular instants. Since in this limit 
$a^3 \sim (t-t_{end})^{2/(1+w)}$, the integrand $\propto a^3 \rho \sim (t-t_{end})^{-2w/(1+w)}$ will not diverge 
provided $|w| < 1$. For $w=+1$ the divergence is at most logarithmic, with coefficients $\simeq {\cal O}(1)$ so that when properly cut off at the physical Planckian density surfaces  these contributions are still much smaller than the cutoff.
Hence our historic averages will always be finite in a bang/crunch universe for all realistic matter sources. Having shown that (\ref{ints}) is bounded, we can proceed with a more careful comparison of contributions to (\ref{ints}) from different cosmological epochs.

Now, to get a more accurate estimate of various contributions to (\ref{ints}) we can split the history of the universe into epochs governed by different 
matter sources, which we can approximate as perfect fluids. So let us consider one such epoch, during which a fluid with equation of state 
$w_i$ controls the evolution of $a$ in an interval $a_i < a< a_{i+1}$. The contribution of this epoch to the temporal integral in (\ref{ints1}) is
\be
I_i =  \int_{t_i}^{t_{i+1}}  dt a^3= \int_{a_i}^{a_{i+1}}  da \, \frac{a^2}{H} =\frac{a_{i+1}^3}{H_{i+1} }\int_{a_i}^{a_{i+1}} \frac{da}{a_{i+1}} \left(\frac{a}{a_{i+1}}\right)^2 \frac{H_{i+1}}{H} \, ,
\ee
where $H=\dot a /a$ and $H_i, ~H_{i+1}$ are the values of $H$ at $a=a_i, ~a_{i+1}$ respectively. Using the Friedmann equation, 
$H^2 \simeq H_{i+1}^2\left(\frac{a_{i+1}}{a}\right)^{3(1+w_i)}$,  and so
\be
I_i= \frac{2}{3(3+w_i)} \left( \frac{a_{i+1}^3}{H_{i+1} }\right)\left[1-\left(\frac{a_i}{a_{i+1}}\right)^{3(3+w_i)/2}\right] \, .
\ee
For $|p/\rho| \le 1$, this integral is always finite, even if $a_i \to 0$. Similarly, the contribution of this epoch to the temporal integral in (\ref{ints}) is 
\ba
J_i &=&  \int_{t_i}^{t_{i+1}}  dt a^3(-\rho+3p) = (3w_i-1) \frac{\rho_{i+1} a_{i+1}^3}{H_{i+1} }\int_{a_i}^{a_{i+1}} \frac{da}{a_{i+1}} \left(\frac{a}{a_{i+1}}\right)^2 \left( \frac{H_{i+1}}{H}\right) \frac{\rho}{\rho_{i+1}}\nonumber \\
& =& (3w_i-1) \frac{\rho_{i+1} a_{i+1}^3}{H_{i+1} }\int_{a_i}^{a_{i+1}} \frac{da}{a_{i+1}} \left(\frac{a}{a_{i+1}}\right)^2 \left( \frac{H}{H_{i+1}}\right) \, ,
\label{jint1}
\ea
where in the last step we have used $H^2 \simeq H_{i+1}^2 \frac{\rho}{\rho_{i+1}}$. 
Since  $H^2\simeq H_{i+1}^2\left(\frac{a_{i+1}}{a}\right)^{3(1+w_i)}$, 
\be
J_i= \frac{2(3w_i-1)}{3(1-w_i)} \left( \frac{\rho_{i+1} a_{i+1}^3}{H_{i+1} }\right)\left[1-\left(\frac{a_i}{a_{i+1}}\right)^{3(1-w_i)/2}\right] \, .
\label{jint}
\ee
Clearly these integrals describe contributions from both expanding and contracting regimes.  

Note, that (\ref{jint1}) is logarithmically divergent near the singularity for a stiff fluid $w_i=1$, as we noted above. But this divergence is a red herring. We cut the evolution off
at a time when the density reaches Planck scale, $l_{pl} \la a \la a_\star$,  and so 
\be
J_\text{stiff} ^\text{singularity} =2   \left( \frac{\rho_{\star} a_{\star}^3}{H_{\star} }\right) \log (\frac{a_\star}{l_{pl}}) \simeq  
\left( \frac{H_{\star} }{M_{Pl}}\right) \log (\frac{a_\star}{ l_{pl}}) \, ,  
\ee
where $H_\star$ and $\rho_\star$ are the Hubble scale and energy density when $a=a_\star$. Hence $J_\text{stiff} ^\text{singularity} \le 1$.  
For all other cases with $-1 \le w <1$, the integral $J_i$ is automatically finite -- and small. 

Next we consider the contributions from the turning point at a time $T$. During this time, the universe is approximately static, with the scale factor roughly a constant, $a \simeq a_{max}$, over a time interval $\Delta t$. The turning point happens roughly at the time given by the total age of the universe, $T \simeq \Delta t \simeq 1/H_{age}$. The effective Hubble parameter at that time is approximately zero, by virtue of a cancellation between different contributions to the energy density, where some must be negative to trigger the collapse (eg., a positive spatial curvature or a negative potential). But since $1/a^2_{max}$ measures the characteristic curvature of the universe at that time, $|R| \sim H_{age}^2  \sim 1/a^2_{max}$, we obtain 
\be
I_\text{turn}=\int^{T+\Delta t/2}_{T-\Delta t /2} dt a^3\simeq a_{max}^3 \Delta t \sim \frac{a_{max}^3}{H_{age}}  \sim \frac{1}{H_{age}^{4}} \, .
\label{inttruni}
\ee
The contribution from the turning point to the temporal integral in (\ref{ints}) is 
\be
J_\text{turn}=\int^{T+\Delta t/2}_{T-\Delta t /2} dt a^3(-\rho+3p) \simeq {\cal O}(1) a_{max}^3 \rho_{age} \Delta t \sim \frac{\rho_{age}}{H_{age}^4} \, ,
\label{inttrunj}
\ee
where $\rho_{age}  \sim M_{Pl}^2 H_{age}^2$ corresponds to the characteristic energy density of the universe at that time.

Clearly, (\ref{inttruni}) and (\ref{inttrunj}) are dominant contributions to (\ref{ints1}) and (\ref{ints}), respectively. To see this, consider first two consecutive epochs away from the turning point. After simple algebra one finds ${I_{i}}/{I_{i-1}}= {\cal O}(1) \left( \frac{a_{i+1}}{a_i} \right)^{3(3+w_i)/2}$, ${J_{i}}/{J_{i-1}} = {\cal O}(1) \left( \frac{a_{i+1}}{a_i} \right)^{3(1-w_i)/2}$ using the Friedmann equation. Clearly, the epoch with larger value of the scale factor at any of its end points gives a dominant contribution. This means that the larger contributions to (\ref{ints1}) and (\ref{ints}) come from the regimes of evolution nearer the turning point. Indeed, at the turnaround, we find that the contributions from the quasi-static interval at during the turning and the adjacent epochs are ${I_\text{turn}}/{I_{i}} = {\cal O}(1) \left( \frac{H_{i+1} } {H_{age} }\right)\left( \frac{a_{max}^3}{ a_{i+1}^3  }\right)$ and 
${J_\text{turn}}/{J_{i}}= {\cal O}(1) \left( \frac{H_{age} } {H_{i+1} }\right)\left( \frac{a_{max}^3}{ a_{i+1}^3  }\right)$. Now 
$H_{age}<H_{i+1}$, and $a_{max}>a_{i+1}$. Next, as long as $|w_i| \le 1$, we also have $H^2_{age} \ga H_{i+1}^2 \left(\frac{a_{i+1}}{a_{max}}\right)^6$, which follows from the fact that the stiff fluid yields the fastest allowed decrease of $H$ with expansion. From these inequalities we conclude that the contributions to (\ref{ints1}) and (\ref{ints}) are indeed dominated by the evolution near the turning point. Thus we obtain 
\ba
\left[\int d^4 x \sqrt{g}\right]_{FRW} &=&{\cal O}(1) \frac{\text{Vol}_3}{H_{age}^4}  \, ,
\label{intres1} \\
\left[\int d^4 x \sqrt{g}\tau^\alpha{}_\alpha\right]_{FRW} &=&{\cal O}(1)  \frac{\text{Vol}_3 \, \rho_{age}}{H_{age}^4} \, .
\label{intres}
\ea

Therefore the residual cosmological constant is 
\be
\Lambda_\text{eff} =\frac14 \langle \tau^\alpha_\alpha \rangle \simeq {\cal O}(1) \rho_{age} \simeq {\cal O}(1) M_{Pl}^2 H_{age}^2  \la M_{Pl}^2 H_0^2 \, .
\ee
This  is guaranteed to be bounded by the current critical density of the universe as long as the universe lives at least $H_{age}^{-1} \ga H_0^{-1} \sim 10^{10}$ years. This proves the previously stated claim that $\Lambda_\textrm{eff}$ is automatically small enough in large old universes, which is a crucial check of our proposal. On the other hand, this can't -- by itself -- be taken as a prediction of the late epoch of cosmic acceleration. For one, the sign of $\Lambda_\textrm{eff}$ is determined by equation of state of the dominant fluid close to the turning point; if $w>1/3$, 
$\Lambda_\textrm{eff}$ is positive, and if $w<1/3$ it is negative. Secondly, if the residual cosmological constant were dominant today, future collapse would have been impossible as long as all other matter satisfies standard energy conditions. Thus $\Lambda_\textrm{eff}$ cannot be identified with dark energy, since the universe must collapse in order to be phenomenologically acceptable within the context of our proposal. The current cosmological acceleration must be a transient phenomenon, with the net potential turning negative some time in the future, 
and/or our universe were spatially closed, with a small but nonzero positive spatial curvature. We will return to this issue in the future \cite{KP3}.

\subsection{Historic integrals and black holes} \label{sec:bh}

One may worry\footnote{We thank Paul Saffin and Alex Vikman for raising this question.} that astrophysical black holes may provide a significant contribution to the spacetime volume of the universe. The point is that the familiar black hole solutions in GR have infinite volumes in their interiors.
However, black holes in compact spacetimes are different. For simplicity let us model a small black hole in a collapsing universe as a Schwarzschild black hole on a spacetime where time is an interval limited by the total age of the universe, $\Delta t \la 1/H_{age}$. The analytic extension across 
the horizon then trades the radial and time coordinates, making $t$ spacelike inside. But because $t$ is compactified, that automatically means the internal volume is finite. Further reduction of black hole contributions may come from the fact that they evaporate, although this seems to be much less important for astrophysical black holes which mainly gain weight in the course of their lifetime. Properties of black holes in compactified spacetimes have been studied in \cite{barak}.

Now we estimate the black hole internal volume. We will stick with the model of a small Schwarzschild black hole with a compactified time direction.
The geometry is approximately $ds^2=-(1-r_H/r) dt^2+ ({1-r_H/r})^{-1} dr^2 +r^2 d \Omega_2$, where $d\Omega^2$ is the metric on the unit $2$-sphere, and $r_H$ is the Schwarszchild radius. So the interior spacetime volume is 
\be
\int _{interior}d^4 x \sqrt{g} =4 \pi \int_0^{\Delta t} dt \int_0^{r_H} dr  r^2=\frac{4 \pi }{3} r_H^3 t \la \frac{4\pi}{3} r_H^3/H_{age} \, .
\ee
The total contribution from all black holes to the spacetime volume of the universe cannot  significantly exceed the contribution from the largest black holes known to exist, with a mass $\sim 10^9 M_{\odot}$ \cite{Vika}. Assuming there is one such black hole in all $10^{11}$ galaxies in the Hubble volume  today,  and assuming that our Hubble volume is typical, we extrapolate the galaxy population in the whole collapsing universe to be about
$N_{gal} \sim 10^{11} H_0^3 a_0^3$, where $H_0 \sim 10^{-33}$ eV is the current Hubble scale and $a_0>10/H_0$ the current scale factor. This yields 
\be
\left[\int d^4 x \sqrt{g}\right]_{galBH}  \la 10^{11} \frac{H_0^3a_0^3 }{H_{age}} \left(10^9 \frac{M_\odot}{M_{Pl}^2}\right)^3 =10^{-19}  \left(\frac{a_0}{a_{max}} \right)^3   \frac{1}{H_{age}^{4}} \, ,
\ee
as the contribution of all black hole interiors to the total spacetime volume of the universe.
We have used that the scale factor near the turning point is $a_{max} \sim 1/H_{age}$. Since $a_0< a_{max}$, we see that galactic black hole interiors' contribution to the spacetime volume is completely negligibe in comparison to the background cosmology. 

Estimating the contribution of their mass to $\langle \tau^\alpha{}_\alpha \rangle$ is now straightforward. Since the total volume integral is unaffected, and since $\int _{interior}d^4 x \sqrt{g} \tau^\alpha{}_{\alpha}  \la M_{BH} \Delta t \la M_{BH}/H_{age}$, where $M_{BH}$ is the total mass in all black holes, the black hole contribution to $\langle \tau^\alpha{}_\alpha \rangle$ is bounded by $M_{BH} H_{age}^3$. Again using the extrapolated number of black holes in the whole universe to be $N_{gal} \sim 10^{11} H_0^3 a_0^3$, the total mass in black holes is
$M_{BH} \sim 10^{20} H_0^3 a_0^3 M_{\odot}$. So after a straightforward calculation, using $M_\odot \simeq 10^{39}  M_{Pl}$, we find that the contribution to $\langle \tau^\alpha{}_\alpha \rangle$ is bounded by
\be
\langle \tau^\alpha{}_\alpha \rangle_{galBH} \la  \frac{\rho_{now}}{10}  \left(\frac{a_0}{a_{max}} \right)^3 \, ,
\ee
where $\rho_{now}=M_{Pl}^2 H_0^2$ is the critical density today. As $a_0 \la a_{max}$, this is clearly subleading even with our overestimate of the black hole population, 
$\rho_{BH} \sim \frac{M_{BH}}{a_0^3} \sim 0.1 \rho_{now}$. 

\subsection{Historic integrals and FRW cosmology} \label{sec:ex}

So far we have been considering the implications of the historic integrals and averages on the phenomenology of solutions, assuming they exist.
Do they? Now we address this question\footnote{We thank Guido D'Amico and Matt Kleban for discussions on the subject of this section.}. In principle, one should expect that such solutions should exist, on the grounds that one could always start with a family of solutions in GR with an arbitrary value of cosmological constant, construct the geometry, and then pick the specific value of the -- initially arbitrary -- bare cosmological constant to satisfy 
$\Lambda_\textrm{eff} = \langle \tau ^\alpha{}_\alpha \rangle/4$. Equivalently, this merely means taking the special solution from each one parameter family of solutions, parametrized by $\Lambda_{bare}$, for which $\langle R \rangle=0$. In a way, the freedom of picking $\Lambda$ and the arbitrariness of initial conditions appear to guarantee the existence of specific solutions. Nevertheless, an explicit proof is still required given that the determination of
$\Lambda_\textrm{eff} = \langle \tau ^\alpha{}_\alpha \rangle/4$ involves specific future boundary conditions which must be picked to find the self-consistent solution. This `bootstrapping' logic is very similar to what one encounters in BCS theory of superconductivity, where one also has to solve a nonlocal equation -- the gap equation -- to decide if the solutions exist in the first place. Here, we will review the conditions under which such solutions do exist in the family of FRW cosmologies. In a forthcoming publication \cite{KP3} we will consider the requirements to build a fully realistic model, consistent with cosmic phenomenology, that includes a transient phase of acceleration like the one we see today. 

For simplicity we will focus on solutions which are dominated by a single fluid and the residual effective cosmological constant. Adding more ingredients in fact makes the existence of solutions easier to prove, by adding additional parameters. So, 
FRW cosmologies with spatial curvature $k$, a cosmological constant $\Lambda_\textrm{eff}$ and a single perfect fluid with equation of state parameter
$w = p/\rho = {\rm const}$, which obeys DEC and NEC, $|w|\le 1$, are described by the Friedmann equation 
\be
3M_{Pl}^2 \left(H^2 + \frac{k}{a^2}\right) = \rho_0 \Bigl(\frac{a_0}{a}\Bigr)^{3(1+w)}  + \Lambda_\textrm{eff} \, .
\label{frw}
\ee
The special solution of (\ref{frw}) which also satisfies Eq. (\ref{residual}), which in this case reduces to
\be
 \Lambda_\textrm{eff} = -\frac{(1-3w)}{4} \frac{\int dt a^3 \rho}{\int dt a^3} \, ,
 \label{lamsign}
 \ee
is the solution of our theory. We will assume that the spatial volume is compactified, so that the integral over the spatial coordinates stays finite.
We will further assume that $\rho_0 > 0$. It is instructive to rewrite
(\ref{frw}) in the form resembling the energy conservation equation for a particle in one dimension,
\be
\dot a^2 + V_\textrm{eff}(a) = - k \, , \qquad V_\textrm{eff} = - \frac{\kappa^2}{a^{1+3w}}  -  \Omega^2 a^2 \, ,
\label{frw2}
\ee
where $\kappa^2 = \frac{\rho_0 a_0^{3(1+w)}}{3 M_{Pl}^2} >0$, but $\Omega^2 = \frac{\Lambda_\textrm{eff}}{3M_{Pl}^2}$ can take either sign. 
 
Let us first consider the (simpler) case of universes with $k \le 0$. 
When $w\ge 1/3$, by (\ref{lamsign}) $\Lambda_\textrm{eff}$ and $\Omega^2$ are nonnegative. So the `potential' $V_\textrm{eff}$ in (\ref{frw2}) is negative definite.
For the total conserved `energy' $-k \ge 0$, the solutions exist for all $\dot a \ne 0$, since the lines $V_\textrm{eff}$ and $\dot a^2 = const > 0$ never intersect. An expanding solution expands forever. Therefore the temporal `volume' $\int dt a^3$ is infinite, and so $\Lambda_\textrm{eff}$ vanishes. However, this also means that $\int d^4 x\sqrt{g}$ diverges. Hence these solutions are all cosmologies in which field theory has no mass gap, and are not good candidates to accommodate our universe. 

When $-1<w<1/3$, $\Lambda_\textrm{eff}$ and $\Omega^2$ are negative. The potential $V_\textrm{eff}$ is a sum of two powers,
$a^{-(1+3w)}$ and $a^2$, with opposite coefficients. However, since $w>-1$, the quadratic always wins at large $a$. This means that a `particle' moving from the origin (an expanding universe starting with a Bang)  with a total energy which is nonnegative ($-k\ge 0$) encounters a potential barrier at some finite $a$ from the origin and turns around. So such configurations always admit a collapsing solution, and for the given `free parameters' describing the solutions of (\ref{frw}) in GR ($\Lambda_\textrm{eff}$ and the value of $a$ at the turning point, specified by the integration constant) one needs to pick the combination which solves (\ref{lamsign}), which as we see exists. Note, that this does not mean tuning the initial conditions for $a$ to find the solution. It means, for the given initial conditions specifying the solution, one needs to pick the right value of the {\it aposteriori} parameter $\Lambda_\textrm{eff}$. The same logic applies to all cases.

The case $k > 0$ is slightly more subtle. In the language of the one dimensional dynamics (\ref{frw2}) we are now looking for states with negative conserved energies. Now, when $-1<w<-1/3$, since $\Lambda_\textrm{eff}$ and $\Omega^2$ are negative, and so is $1+3w$, the potential $V_\textrm{eff}$ is a sum of two positive powers of $a$, with opposite coefficients. This sum is negative between the origin and some (large) value of $a$, beyond which it turns positive, going again as $a^2$ at large $a$. But since we are looking for trajectories with $-k<0$ now, it means that the total energy $-k$ can be greater than the effective potential $V_\textrm{eff}$ only in a finite interval of $a$'s. So classical motion is only possible between these two turning points, and it continues forever. This case corresponds to oscillating cosmologies of \cite{rajpoot}, which have finite spatial volume, and where the negative effective cosmological constant, needed for oscillating motion, is generated by our constraint. Here,
even though $\int d^4 x \sqrt{g}$ is infinite, and so the QFT is gapless, $\Lambda_\textrm{eff}$ remains finite as can be seen readily by splitting the integration into the sum of integrals over full periods. Yet, such universes are not phenomenologically viable since QFT is scale invariant.

When $-1/3 < w <1/3$, $\Lambda_\textrm{eff}$ and $\Omega^2$ are still negative, but $1+3w > 0$. Thus $V_\textrm{eff}$ is a combination of an infinite potential well at the origin and a quadratic barrier far away. So solutions with $-k <0$ always exist, again representing universes which start with a Bang, expand to the maximum radius and subsequently crunch. The cases $w=-1/3$ and $w=1/3$ are special limits. In the former, the barrier has finite depth, again admitting collapsing solutions, while in the later case the barrier is pushed to infinity since $\Lambda_\textrm{eff}=0$ by scale invariance. This latter case corresponds to a radiation dominated universe with vanishing cosmological constant. 

Finally, when $w>1/3$, $\Lambda_\textrm{eff}$ and $\Omega^2$ are positive. Hence the potential $V_\textrm{eff}$ is negative definite, diverging at the origin and infinity, with a maximum in between. Since $-k<0$, solutions exist if the `conserved energy' $-k$ is smaller than the value of $V_\textrm{eff}$ at the maximum, representing again cosmologies that start with a Bang and end with a Crunch. The solutions would not have existed if $-k$ were larger than the maximum. But this is not the case; the limiting case, where $-k$ is exactly equal to $V_\textrm{eff}(max)$, would have been a static Einstein universe, that would have been eternal. This would require $\ddot a = \dot a = 0$. Taking the derivative of (\ref{frw2}) one can easily check that this requires $w=-1$, contradicting $w>1/3$. So therefore the closed cosmologies describing Bang/Crunch always exist for $w>1/3$.

Before closing this section, we should clarify the role and the implications of the constraint $\langle R \rangle = 0$ which follows from tracing and integrating Eq. (\ref{eeqs1}) and using 
Eq. (\ref{residual}). On FRW geometries with spatially compact smooth sections it reduces to
$\int dt a^3 (\dot H + 2 H^2 + \frac{k}{a^2}) = 0$ after factoring out the finite spatial integral.
Integrating the term $\propto \dot H$ by parts yields
\be
a^3 H\Bigl|^{t_{crunch}}_{t_{bang}} = \int dt a^3 (H^2 - \frac{k}{a^2})  \, .
\label{frwconstr}
\ee
If we were to take the limits of integration to be the time intervals where the scale factor literally vanishes, and the dominant stress-energy sources determining the behavior of $H$ obey NEC, the left-hand side would have vanished for precisely the same reasons as we have already discussed in Sec. \ref{sec:histint}. This would appear to have ruled out spatially open and flat FRW universes, as is clear from Eq. (\ref{frwconstr}) since in this case the right hand side could only have vanished if $H$ and $k$ were exactly zero at all times. 
This argument is a little hasty, however. The point is that the integration must be ended at times when the curvature $R$ reaches the Planck scale, and not when the scale factor exactly vanishes. Thus the left hand side does not vanish, but simply represents a boundary condition which the terminal geometry must satisfy to ensure that 1) it is a solution of Einstein's equations and 2) that the vacuum energy is sequestered. Indeed, the left hand side is merely a nonzero number which represents the difference between $a^3 H$ at the beginning and the end of cosmology, encoding the posterior determination of the effective cosmological constant $\Lambda_{eff} = \langle \tau \rangle/4$, communicated to the on-shell geometry by way of Einstein's equations. For small universes, the contributions to this term from beyond the cut-off are sufficiently important that we can treat it as an
arbitrary boundary condition allowing any value of  $k$. However: for large and old universes, the integrals on the right hand side are dominated by the contributions from large volumes near the turning point, and thus essentially insensitive to the boundary conditions. Since the left hand side is much smaller, the integrals on the right hand side must include negative contributions that yield cancellations to match the left hand side. Thus having a large old approximately FRW universe
does require $k =1$ (and so $\Omega_k < 1$). Small inhomogenous universes may evade this. A similar observation was made in a different context in \cite{barrow}.

\section{Vacuum energy sequestering and phase transitions} \label{sec:phase}

In the course of the cosmological evolution the field theory dynamics may go through many phase transitions: QCD phase transition, 
electroweak symmetry breaking, etc. This means that the local vacuum of the theory changes. As a result, since after renormalization the effective potential governing the dynamics is fixed, this implies that the net renormalized cosmological constant may actually change in time and space. 
The variation is typically of the order of $\Delta V \sim M_{PT}^4$, where $M_{PT}$ is the energy scale of the transition. So the question arises, if the universe 
has undergone many phase transitions in its past, how come that the initial finite value of cosmological constant was just right to exactly cancel the phase transitions and yield a very small value which obeys current observational bounds \cite{dreitlein,linde,veltman}?

In our framework, such contributions do {\it not} drop out from (\ref{eeqs}), (\ref{eeqs1}), but they become {\it automatically} small
at times after the transition in a large and old universe.  Let us assume, for simplicity, a single phase transition that took place in the past.  
If the phase transition is first order, it proceeds through bubble formation, which will percolate some time after the transition. If it is second order, then it is characterized by some order parameter smoothly transitioning from one value to another in the universe, where the onset of this rolling in different regimes is local. In any case the process takes some time to complete, and therefore involves gradients of fields or background geometry. Either way, the energy density in the gradients will not exceed the total potential energy density change. So for simplicity we will model the phase transitions as a sudden jump in the potential across the whole universe. This means that we ignore the gradients of fields, as they just give an ${\cal O}(1)$ correction,  and model the transition with a simple step function potential   $V = V_{before} (1- \Theta(t-t_{*}))+ V_{after} \Theta(t-t_{*})$, where $ \Theta(t-t_{*})$ is the step function, and $t_{*}$ the transition time. Substituting into  (\ref{eeqs1}),  we find
\be
\tau^\mu{}_\nu-\frac{1}{4} \delta^\mu{}_\nu \langle  \tau^\alpha{}_\alpha \rangle=
\begin{cases} -\langle V_{before}-V \rangle  \delta^\mu{}_\nu \qquad & t<t_*  \, ,\\
-\langle V_{after}-V \rangle \delta^\mu{}_\nu \qquad & t>t_* \, . \end{cases}
\label{vtran}
\ee
Hence, after the transition, the historic average in (\ref{vtran}) is
\be \label{trans}
\langle V_{after}-V \rangle=-\Delta V \frac{\int_{t_{bang}}^{t_{*}} dt a^3 }{\int_{t_{bang}}^{t_{crunch}} dt a^3} \, ,
\ee
where  $\Delta V = V_{before} - V_{after}$. The denominator is just the spacetime volume, computed in section \ref{sec:histint}, and is given by $\sim 1/H_{age}^4$. To estimate the numerator, we use the result of section \ref{sec:histint}, where we showed that 
the largest contributions to the temporal integrals during a particular epoch come from the region where the scale factor is largest. 
Since we are considering the phase transitions in the past, as dictates by  the structure of the Standard Model, the dominant contributions come from just before the transition itself. Hence
$$
\int_{t_{bang}}^{t_{*} } dt a^3 \sim {\cal O}(1) \frac{a_*^3}{H_*} \, ,
$$
where $a_*$ is the scale factor and  $H_*$ the curvature scale during the transition. Therefore
\be
\langle V_{after}-V \rangle={\cal O}(1)  \rho_{age} \frac{\Delta V}{M_{Pl}^2 H_*^2}  \left(\frac{H_*}{H_{age}} \right) \left(\frac{a_*}{a_{max}}\right)^3  \, , 
\ee
where we have used the fact that $\rho_{age} \sim M_{Pl}^2 H_{age}^2$, and $a_{max} \sim 1/H_{age}$. Assuming that the cosmological evolution after the transition to the turning point is dominated by a single perfect fluid with an equation of state $w$, so that 
$H_{age}^2=H_{*}^2\left(\frac{a_{*}}{a_{max}}\right)^{3(1+w)}$ we then find that
\be
\langle V_{after}-V \rangle=  {\cal O}(1) \rho_{age} \frac{\Delta V}{M_{Pl}^2 H_*^2}   \left(\frac{H_{age}}{H_{*}} \right)^{\frac{1-w}{1+w}} \, .
\ee
If the phase transition is followed by several different epochs dominated by different matter distributions, the correction is merely a factor of ${\cal O}(1)$. Now assuming DEC and NEC, $|w| \le 1$, along with the fact that  
$H_*  \sim \sqrt{V_{before}}/M_{Pl}$ $\ga \sqrt{\Delta V}/M_{Pl}$, we conclude that
\be
| \langle V_{after}-V \rangle | \la \rho_{age} < \rho_{now} \, ,
\ee
which means that the contributions of phase transitions to the net effective vacuum energy are automatically suppressed to below the critical energy density of the universe today. As we also see, the earlier they are, the more suppressed: the reason is the historic weighting of the contributions, since the `imbalance' is weighted only by the spacetime volume before the transition. Dividing by the total spacetime volume of the universe then suppresses the contributions from the transition more efficiently if they occur earlier. For the Standard Model all the phase transitions occur early, with $H_* \gg H_0> H_{age}$, and so the suppression is considerable. 

What about the period before the transition? Using similar arguments, we find that $\langle V_{before}-V \rangle   \sim {\cal O}(1) \Delta V$.  Since $\Delta V \sim M_{pl}^2 H_*^2$,  this contribution could be large, but it affects the universe only at early times before the transition, when $H \ga H_*$. This means that this {\it aposteriori} contribution is subdominant to the contributions from other energy sources, except at most during the period immediately before the transition. Indeed, it is consistent with the Standard Model and a reheating scale $\gg $TeV to assume that all transitions occur during the radiation era. So this could at most yield several short bursts of inflation (of at most a few e-folds) in the run up to a transition, as long as it is sharp. Such phases are only a small perturbation of the radiation driven cosmology, and occurring at very short scales in the present universe. 
If, however, the transition is very slow, say a second order phase transition where the order parameter takes a long time to move from its initial position to the new vacuum, this in fact can drive a long inflation responsible for making the universe subsequently big and old. This, in fact, is beneficial, and we will show in section \ref{sec:inflation} that it is the key reason why our proposal is consistent with inflation, even in the protected sector.

\section{Particle masses and phenomenology} \label{sec:masses}

So far we have shown that vacuum energy can be sequestered away from curvature yielding a small residual net cosmological constant, while particle physics still has a mass gap as long as the universe is compact in spacetime, with a finite spacetime volume.
Here we expand on the latter point, and discuss in greater detail the properties that the function $\sigma$ must have in order to ensure that the particle mass gap is 
phenomenologically reasonable. Since the numerical value of the parameter $\lambda$ sets the physical scales in ${\cal L}$, setting $m_{phys} = \lambda m$, where $m$ is the bare mass, the function $\sigma$ must be carefully engineered to generate the right hierarchies between the various scales in a given model of (beyond) the Standard Model physics. This {\it cannot} protect the hierarchy between $m_{phys}$ and $M_{Pl}$, nor the hierarchies between different physical masses in ${\cal L}$, but it can help set the scale hierarchy, and it can coexist with specific mechanisms designed by model builders to protect particle physics hierarchies, by removing the vacuum energy contributions.

Since by virtue of the equations (\ref{varsl})
\be
\frac{m_{phys}}{m}=\lambda=\left[\frac{\sigma'\left(\langle \tilde T^\alpha{}_\alpha\rangle /4 \mu^4  \right)}{\mu^4 \int \sqrt{g} }\right]^{1/4} \, ,
\ee
to compute it we need to specify the cutoff for the regulated vacuum energy contributions. We will assume that the cutoff is large enough that the vacuum energy gives the dominant contribution to $\tilde T^\alpha{}_\alpha$. We approximate $\tilde T^\alpha{}_\alpha \approx -4 \tilde V_{vac}$, where $\tilde V_{vac} =\langle 0| {\cal L}_\textrm{eff} (\tilde g^{\mu\nu}, \Phi |0\rangle $ is the regulated vacuum energy of the effective matter Lagrangian in the protected sector ${\cal L}$ at any given order in the loop expansion. It is related to the {\it physical} regulated vacuum energy via the relation $V_{vac}=\lambda^4 \tilde V_{vac}$.  The regulator scale, ie., the cut-off, of the protected sector is $M_{UV}^{phys} =\lambda M_{UV}$, and in principle it can be as  high as  $M_{Pl}$. Depending on the model, the `bare' regulator, $M_{UV}$, could also be as high as $M_{Pl}$, in which case we must ensure that the solutions saturate $\lambda  \sim {\cal O}(1)$.  Even in this most extreme scenario, the regulated vacuum energy $V_{vac} \sim \tilde V_{vac} \sim M_{Pl}^4$, is still  completely sequestered from gravity by our mechanism. The scale $\mu$ needs to be chosen appropriately in order to be consistent with $\lambda  \sim {\cal O}(1)$, via the condition
\be
{\cal O}(1)=\left[\frac{\sigma'\left( \pm {\cal O}(1) M_{Pl}^4/ \mu^4  \right)}{\mu^4 \int \sqrt{g} }\right]^{1/4} \sim \left[\sigma'\left( \pm {\cal O}(1) M_{Pl}^4/ \mu^4  \right)\right]^{1/4} \frac{H_{age}}{\mu} \, ,
\ee
where the $\pm$ reflects the fact that the vacuum energy may take either sign. Since we are introducing $\mu$ buy hand, we can take it to have as natural a value as possible, and pick it close to the cut-off in the protected sector. Note, that this parameter is entirely {\it external} - a mere normalization required for dimensional purposes - since neither it nor the function $\sigma$ ever get explicitly renormalized. The value of the function $\sigma$ can only change because the ratio $\Lambda/\lambda^4$ which appears in the argument of $\sigma$ may change between different orders in the loop expansion. So setting $\mu \sim M_{Pl}/\sqrt{10}$, we now require
\be
{\cal O}(1)\sim  \left[\sigma'\left( \pm {\cal O}(100)  \right)\right]^{1/4} \frac{H_{age}}{M_{Pl}} \, .
\ee
Provided the lifetime of the universe, $1/H_{age}$, does not exceed its lower bound, $10/H_0$, by too much, getting $\lambda \sim {\cal O}(1)$ is will follow if $\sigma(z) \simeq \exp(z)$ for $z>1$. To allow for both positive and negative vacuum energy, we need an odd function, so we can simply take $\sigma(z) \sim \sinh z$. 

We should note that the exponential form of $\sigma$, which easily sets up the required hierarchy between the electroweak scale and the Planck scale is also highly beneficial by protecting particle physics masses from a significant sensitivity to the cosmological initial conditions. This can be easily seen as follows. Consider Eqs. (\ref{varsl}) and imagine, for example, that $\sigma(z) = z$. In this case, the first of (\ref{varsl}) translates to 
$\lambda =1/[\mu (\int d^4x \sqrt{g})^{1/4}]$, which by Eq. (\ref{intres1}) implies $\lambda \simeq H_{age}/\mu$. In this case, if we take the cutoff to be $M_{UV} \sim M_{Pl}$, to get $m_{phys} \sim {\rm TeV}$, we need $\lambda \sim 10^{-15}$. If $H_{age} \sim H_0 \sim 10^{-33} {\rm eV}$, this requires $\mu \sim 10^{-18} {\rm eV}$. Besides having such a tiny $\mu$, this setup is even more problematic from extreme sensitivity to the cosmological conditions that yield the terminal value of $H_{age}$. For example, if they are changed so that one has a few more or less efolds of inflation, $H_{age}$ will change by orders of magnitude. If this happens, one must redial all the particle scales by those amounts by hand. Having an exponential form of $\sigma$ reduces this sensitivity to logarithms, which are far tamer.

Of course, the Standard Model may be embedded in ${\cal L}$ via some of its BSM extensions, such as some version of supersymmetry, in which case we could get a much lower value of the vacuum energy. For example, the SUSY breaking scale might still be close to TeV, in which case 
$\Lambda_{vacuum} \ga ({\rm TeV})^4$. In this case, to retain $\lambda \sim {\cal O}(1)$ as the solution, and keep the exponential form for $\sigma$, 
which protects particle scales from sensitivity to cosmological initial conditions, we ought to rescale $\mu$ so that it is closer to TeV. 
Either way, in principle we can adjust the external parameters $\mu$ and/or $\sigma(z)$ to fit with a dynamical mechanism which 
protects the hierarchy within ${\cal L}$ . Then our mechanism can complement the QFT hierarchy protection by sequestering those vacuum energy contributions which the BSM extension cannot remove.

\section{Planck mass renormalization and generalized actions} \label{sec:gen}

So far we have conspicuously ignored discussing the renormalization of the Planck mass. In the context of QFT coupled to (semi) classical gravity, however, we cannot avoid facing this issue. The reason is precisely the logic we have adopted in addressing the problem of vacuum energy corrections
to the cosmological constant. These are calculated in the loop expansion in QFT, and involve loops with gravitons only in external lines. However, similar loop diagrams also give rise to graviton vacuum polarization - ie., wave function renormalization. These are precisely the corrections to the Planck mass. For example, if one includes one-loop corrections to the graviton wave function, one finds that each field theory species contributes to 
the regulated Planck mass an expression of the form $\Delta M^2_{Pl} \simeq {\cal O}(1) 
\times ({M}_{UV}^{phys})^2 + {\cal O}(1) \times 
m_{phys}^2 \ln({M}_{UV}^{phys}/m_{phys}) + {\cal O}(1) \times
m_{phys}^2 + \ldots$, where ${M}_{UV}^{phys}=\lambda {M}_{UV} $ is the matter UV regulator 
mass and $m_{phys} = \lambda m_{bare}$ the mass of the virtual particle in the loop \cite{myers}. It is important to note that here, as well as in the corrections to the vacuum energy, the dependence on $\lambda$ is multiplicative, since $\lambda$'s in the logs cancel by virtue of the careful choice of the regulator. Renormalization of $M_{Pl}$ then corresponds to picking a suitable subtraction scale ${\cal M}$ and canceling the $M_{UV}$ dependent pieces, which yields 
$M_{Pl}^2 \to M_{Pl}^2 + {\cal O}(1) \times {\cal M}_{phys}^2 + {\cal O}(1) \times m_{phys}^2 \ln\left(\frac{{\cal M}_{phys}}{m_{phys}}\right) + {\cal O}(1) \times
m_{phys}^2 + \ldots$. But, since ${\cal M}_{phys} = \lambda {\cal M}$, this implies that the counterterm in the renormalization of the Planck scale must also go as $\lambda^2$, implying that the consistent semiclassical theory really is given by the action 
\be
S= \int d^4 x \sqrt{g} \left[ \frac{M^2_{Pl}+\lambda^2 M^2}{2} R  - \Lambda - {\lambda^4} {\cal L}(\lambda^{-2} g^{\mu\nu} , \Phi) \right] +\sigma\left(\frac{ \Lambda}{\lambda^4 \mu^4}\right) 
 \, . 
 \label{action1}
\ee
Here, $\lambda M$ is the total finite renormalization of $M_{Pl}$ which remains after subtracting the infinities.
As long as  ${M}_{UV}^{phys} = \lambda M_{UV} < M_{Pl}$, a condition automatically satisfied by picking a cutoff of the QFT below $M_{Pl}$ and maintained by evolution in a large old universe\footnote{This is because the physical masses scale as $\lambda$, and $\lambda$ decreases with increasing space-time volume.}, the value of the Planck scale is  radiatively stable in our original model. The change of the form of (\ref{action1}) relative to (\ref{action}) then does not affect the dynamics of vacuum energy sequestering at all in the limit of (semi) classical gravity.

To see this, note that the field equations that follow from varying the action (\ref{action1}) with respect to $\Lambda, \lambda$ now take a slightly different form,
\be
\frac{\sigma'}{\lambda^4\mu^4} = \int d^4x \sqrt{g} \, , \qquad \qquad 4\Lambda \frac{ \sigma' }{\lambda^4\mu^4} 
= \int d^4x \sqrt{g} \,\left (\lambda^2 M^2 R+\lambda^4 \, \tilde T^{\mu}{}_\mu\right) \, ,
\label{varsl1}
\ee
yielding
\be
\Lambda = \frac14 \langle \lambda^2 M^2 R+\lambda^4 \, \tilde T^{\mu}{}_\mu \rangle \, .
\label{newleff}
\ee
The variation of (\ref{action1}) with respect to $g_{\mu\nu}$ gives 
$\left(M_{Pl}^2+\lambda^2M^2\right) G^\mu{}_\nu = -\Lambda \delta^\mu{}_\nu + \lambda^4 \tilde T^\mu{}_\nu$. 
After eliminating $\Lambda$ from this equation with the help of Eq. (\ref{newleff}) and canonically normalizing the matter sector, this becomes
\be
\left(M_{Pl}^2+\lambda^2M^2\right) G^\mu{}_\nu = T^\mu{}_\nu-\frac{1}{4} \delta^\mu{}_\nu \langle  \lambda^2 M^2 R+ T^\alpha{}_\alpha \rangle \, .
\label{eeqs*}
\ee
Rewriting this as $M_{Pl}^2G^\mu{}_\nu   = T^\mu{}_\nu-\lambda^2M^2 G^\mu{}_\nu- \delta^\mu{}_\nu \langle T^\alpha{}_\alpha -  \lambda^2 M^2 G^\alpha{}_\alpha\rangle/4$, we see that the right-hand side vanishes identically upon taking the trace and the space-time average. Thus, as long as 
$M_{Pl}^2 \neq 0$, we again obtain $\langle R \rangle=0$, and so the gravitational equations reduce to
\be
\left(M_{Pl}^2+\lambda^2M^2\right) G^\mu{}_\nu = T^\mu{}_\nu-\frac{1}{4} \delta^\mu{}_\nu \langle  T^\alpha{}_\alpha \rangle \, .
\label{eeqs3}
\ee
The theory is identical to the one described previously but with a renormalized Planck mass $(M_{Pl}^\text{ren})^2=M_{Pl}^2+\lambda^2M^2$ and everything we have said so far remains unchanged!

A curious limiting case occurs if we take the bare Planck scale to zero, $M_{Pl}^2 \to 0$, in effect thinking of the whole theory as `induced gravity, 
\cite{sakharov,adler}. The action becomes
\be
S= \int d^4 x \sqrt{g} \left[ \frac{\lambda^2 M^2}{2} R  - \Lambda - {\lambda^4} {\cal L}(\lambda^{-2} g^{\mu\nu} , \Phi) \right] +\sigma\left(\frac{ \Lambda}{\lambda^4 \mu^4}\right) 
 \, .
\label{action2}
\ee
Now the resulting gravitational field equations  have a vanishing historic trace:
\be
0= T^\mu{}_\nu-\lambda^2M^2 G^\mu{}_\nu-\frac{1}{4} \delta^\mu{}_\nu \langle T^\alpha{}_\alpha -  \lambda^2 M^2 G^\alpha{}_\alpha\rangle \, .
\ee
This  degeneracy of the averaged trace is an artifact of the global  scale invariance (\ref{scale}) being unbroken, so that the field equations no longer force the constraint $\langle R\rangle=0$. Nevertheless, the vacuum energy from the protected matter sector is again sequestered from gravity in this model - all loop corrections still automatically cancel. 
Indeed, as above, at any given order in loops, we can separate off vacuum energy and local excitations by writing, $T^{\mu}{}_\nu = V_{vac} \delta^\mu{}_\nu + \tau^\mu{}_\nu$. Then the vacuum energy is seen to drop out, and the field equations become, 
\be
\lambda^2M^2 G^\mu{}_\nu= \tau^\mu{}_\nu-\frac{1}{4} \delta^\mu{}_\nu \langle \tau^\alpha{}_\alpha + \lambda^2 M^2 R\rangle \, .
\label{correoms}
\ee
With the vacuum energy eliminated, the theory now behaves like GR with an effective Planck mass $M^{phys}_{Pl}=\lambda M $, and a residual cosmological constant $\Lambda_\text{eff}=\langle \tau^\alpha{}_\alpha + \lambda^2 M^2 R\rangle/4$. 
Unlike the generic class of theories, in this case the choice of $\Lambda_\text{eff}$ does {\it not} impose any boundary conditions on the matter sources. This of course is a consequence of the fact that the whole theory now has unbroken scale invariance. Choosing $\Lambda_\text{eff}$ breaks it spontaneously, but it does not fix $\langle \tau^\alpha{}_\alpha \rangle$. Thus this limit of our mechanism has more in common with unimodular gravity than the main case, in that it leaves an unspecified integration constant as the effective residual cosmological constant. However, the vacuum energy corrections from the QFT are still completely cancelled, unlike in the standard formulation of unimodular gravity. Thus the residual cosmological constant is completely classical. 

The unbroken scale invariance of the theory (\ref{action2}) makes it tempting to argue that the sequestering of the vacuum energy corrections from gravity may now be extended to include also the corrections involving virtual graviton lines. This, however, does not happen. While some of the loops involving internal graviton lines might indeed cancel, the presence of $R$ in the historic average in (\ref{correoms}) is indicative of the problem. It 
will be renormalized by quantum gravity effects, both quantitatively and by the fact that other higher derivative operators will appear in the action, 
which implies that the radiative stability of the residual cosmological constant will be lost. In general however, allowing gravitons in the loops takes us on a road to the -- as yet unexplored -- realms of quantum gravity, so it is difficult to say anything concrete. To sequester contributions from graviton loops at each and every order we would presumably need stronger symmetry requirements than just global scale invariance which appears here.
Therefore we will refrain from delving into this complex issue here, hoping to return to this in the future.

Note that the phenomenology of this limiting case of our mechanism also differs. In this case $\lambda$ does not set the hierarchy between the  physical matter scales and the Planck mass, since $\frac{m_{phys}}{M^{phys}_{Pl}}=\frac{\lambda m}{ \lambda M}=\frac{m}{M}$. This implies that now the spacetime volume is allowed to be infinite, and in particular it admits Minkowski spacetime as a solution. This immediately raises the specter of Weinberg's {\it no-go} theorem, which we have evaded previously by requiring a collapsing spacetime. In the Minkowski limit, our scale invariant  theory corresponds to the runaway behavior as $\tilde \Phi_0 \to -\infty$ discussed at the end of our review of Weinberg's theorem in the previous section. There we argued that such a runaway also set  all QFT scales to zero, in obvious conflict with the universe we see. The caveat here is that in the case of Weinberg's no-go, the Planck mass were held fixed. That is not the case here. The effective Planck mass experiences exactly the same runaway behavior, dependent on the volume, so that ratios between masses are maintained. So it appears that at least in the limit of classical gravity we can consistently take all bare masses to infinity such that the physical masses remain finite. However, it's not clear this will survive when quantum effects are in fact included on the gravitational side, even in infrared. Again this is a question which we leave to future considerations.

Finally, let us comment here on the main difference between our proposal and the suggestion by Tseytlin \cite{tsey} which seems to be similar (for followups to Tseytlin's work, exploring his proposal at the classical level, see \cite{Davidson}). This proposal posits that the standard action of a QFT minimally coupled to gravity should be divided by the spacetime volume of the universe, $S_T ={\int d^4x \sqrt{g} \left[\frac{M_{Pl}^2}{2} R - {\cal L}(g^{\mu\nu}, \Phi) \right]}/\left[{\mu^4 \int d^4x \sqrt{g}}\right]$. Clearly, this trick immediately removes the classical and zero-point (tree-level) corrections to the cosmological constant. However it {\it does not} remove the loop corrections to vacuum energy. To see that, rewrite the theory as
\be
S _{T} = \int d^4x \sqrt{g} \left[\frac{\lambda^4 M_{Pl}^2}{2} R - \Lambda - \lambda^4 {\cal L}(g^{\mu\nu}, \Phi) \right] + \frac{\Lambda}{\lambda^4 \mu^4} \, ,
\label{tsey}
\ee
with the introduction of the global variables $\Lambda, \lambda$. Ignoring any phenomenological problems associated with the linear fucntion 
$\sigma$, we see that the main difference between (\ref{tsey}) and our proposal (\ref{action}) is the dependence of the bulk terms on $\lambda$. Here,
the Einstein-Hilbert term has a $\lambda^4$ prefactor, and the matter sector does not have the kinetic energy scaling $1/\lambda^2$. Now, it's convenient to normalize the Einstein-Hlbert term canonically,  by taking $g_{\mu\nu} \to \lambda^{-4} g_{\mu\nu}$. The action (\ref{tsey}) then becomes $S _T = \int d^4x \sqrt{g} \left[\frac{M_{Pl}^2}{2} R - \frac{\Lambda}{\lambda^8} -\frac{1}{ \lambda^4} {\cal L}\left(\lambda^{4}g^{\mu\nu}, \Phi\right) \right] + \frac{\Lambda}{\lambda^4 \mu^4}$. Now it is clear that the tree-level vacuum energy scales like $1/\lambda^4$ and so will be automatically eliminated from the dynamics once $\lambda $ is integrated out. However, after canonically normalizing the QFT Lagrangian, it is straightforward see that the physical masses scale as $m_{phys}=m/\lambda^2$, and so the radiative corrections to the vacuum energy scale as $\sim 1/\lambda^8$. Thus they will not automatically cancel, and will restore the vacuum energy radiative instability in much the same way as in GR (or unimodular gravity). In effect, the $\lambda$ dependence of (\ref{tsey}) does not correctly count the engineering dimension of the vacuum energy loop 
corrections. Further, as already alluded in \cite{tsey}, the Planck mass is {\it not} radiatively stable. The reason is that the corrections to it come as
$\Delta M_{Pl}^2 \simeq {\cal O}(1) \times m_{phys}^2 \simeq {\cal O}(1) m^2/\lambda^4$, and so they are extremely large in old and large universes. Therefore in a large and old universe like ours, the Planck scale would receive very large radiative corrections (unless of course the bare particle masses are incredibly small to start with). This is in complete contrast to what happens in our model.

\section{Inflation} \label{sec:inflation}

So far, we have shown that in our framework the residual effective cosmological constant is automatically small in large old universes. How does a universe become large and old? A common approach to answering this question is to resort to the inflationary paradigm, and postulate that at some early epoch the universe was dominated by a transient large vacuum energy which made it very big and smooth quickly. But, we want to get rid 
of vacuum energy here. So how do we reconcile these two requirements? Is our framework compatible with inflation? 
If so, why does our mechanism not sequester away the vacuum energy during inflation?

In \cite{kalpad1} we noted that in principle one could add an extra sector to (\ref{action}) which contains an inflaton, {\it outside} of the protected sector ${\cal L}$. However, since inflation must end and the universe must reheat, this means that the inflaton must couple to the protected sector fields 
in ${\cal L}$.  Then one must worry about quantum cross-contamination between the two sectors and how this may spoil the vacuum energy sequestration. In particular, loops involving inflaton internal lines could spoil the scaling of the vacuum energy corrections with $\lambda$, and so they could end up yielding large corrections to the residual cosmological constant. We noted however that there is an easy way out: since we are already treating gravity (semi) classically, prohibiting internal graviton lines in vacuum energy loops, we could just embed the inflaton in the gravitational sector, and imagine that some -- yet unknown -- mechanism protects the residual vacuum corrections involving virtual inflaton lines in the same way it protects them from corrections involving virtual gravitons. 

Such a model is readily provided by the  original inflation model of Starobinsky \cite{star}. This model has -- until very recently -- been favored by the  data \cite{data}, and even post BICEP2 \cite{bicep}, there may still be variants in play \cite{starpost}. The model can be simply embedded in our framework by extending the action (\ref{action}) by 
 \be
S= \int d^4 x \sqrt{g} \left[ \frac{M^2_{Pl}}{2} R +\beta \, R^2  - \Lambda - {\lambda^4} {\cal L}(\lambda^{-2} g^{\mu\nu} , \Phi) \right] +\sigma\left(\frac{ \Lambda}{\lambda^4 \mu^4}\right) 
 \, ,
\label{staction}
\ee
where $\beta \sim {\cal O}(10^{6})$ is a dimensionless parameter.  Such a large parameter is radiatively stable to the corrections from the loops
of the protected sector fields, and retains its form as long as we pick the UV regulators that couple to $\tilde g_{\mu\nu}$, as before. We can now go to the axial gauge, by first extracting the Starobinsky scalaron $\chi$ by the field redefinition $\bar g_{\mu\nu} = \left(1+\frac{4\beta}{M_{Pl}^2} R\right) g_{\mu\nu}$, 
$\chi = \sqrt{\frac32} M_{Pl} \ln\left(1+\frac{4\beta}{M_{Pl}^2} R\right)$ \cite{kko}. The scalaron has the potential $V = \frac{M_{Pl}^4}{16 \beta} \left[1- \exp\left(-\sqrt{\frac23} \chi/M_{Pl}\right)\right]^2$ and the protected 
matter sector couples to both $\bar g_{\mu\nu}$ and to $\chi$, via
\ba
S &=&   \int d^4 x \sqrt{\bar g} \left[ \frac{M^2_{Pl}}{2} \bar R  - \frac12 (\bar \partial \chi)^2 - V - \Lambda e^{-2\sqrt{\frac23} \frac{\chi}{M_{Pl}}}  -
{\lambda^4} 
e^{-2\sqrt{\frac23} \frac{\chi}{M_{Pl}}} {\cal L}\left(\frac{e^{\sqrt{\frac23} \frac{\chi}{M_{Pl}}}}{\lambda^{2}} \bar g^{\mu\nu} , \Phi\right) \right]  \nonumber \\
&& +
\sigma\left(\frac{ \Lambda}{\lambda^4 \mu^4}\right) \, . 
\label{starac}
\ea
Clearly, the sequestering of vacuum energy in the protected sector goes through unaffected, which follows from the scaling of the protected sector Lagrangian with $\lambda$. The Starobinsky inflation $\chi$ does not spoil it, since coming from the gravitational sector it is also treated purely classically, like the graviton. By a direct variation of (\ref{starac}) one could directly verify this, noting that a substitution 
$T^{\mu}{}_\nu =-V_{vac}\delta^\mu{}_\nu + \tau^\mu{}_\nu$, where $V_{vac}$ is the physical vacuum energy computed up to any given order in loops, and $\tau^\mu{}_\nu$ is the energy momentum tensor describing local on-shell modes, still leads to a complete cancellation of
$V_{vac}$. 

Furthermore, we find that the deviations from the original Starobinksy scenario go like $\langle \tau^\alpha{}_\alpha \rangle_\chi/V$, where the $\chi$-modified average is defined as  $\langle Q \rangle_\chi =\frac{\int d^4x \sqrt{\bar g}  e^{-2\sqrt{\frac23} \frac{\chi}{M_{Pl}}} Q }{\int d^4x \sqrt{\bar g} e^{-2\sqrt{\frac23} \frac{\chi}{M_{Pl}}} }$. Since $\chi \ne 0$ only during inflation, the dominant contribution to the $\chi$-modified averages
comes from the full history of the universe, which means that  
$  \langle \tau^\alpha{}_\alpha \rangle_\chi  \approx  \langle \tau^\alpha{}_\alpha \rangle  \sim \rho_{age}$. This is very small compared to the inflaton potential $V \sim M_{pl}^4/\beta$ during inflation. We see, then, that the dynamics of Starobinsky inflation with sequestering is identical to the standard case to an accuracy of $\sim \rho_{age}/(M_{Pl}^4/\beta) \la 10^{-110} $.

However, if the BICEP2 claim survives the ongoing scrutiny, the inflationary mechanism which shaped our universe may be different from Starobinsky.
In fact, BICEP2 appears to favor large field chaotic inflation \cite{andreichao}, which can be UV completed consistently \cite{eva,flux}.  It turns out that such inflationary models are also consistent with vacuum energy sequestering. This means, the inflaton itself resides in the protected sector - so that the exit from inflation, reheating and all the usual inflationary phenomenology may proceed without spoiling the cancellation of vacuum energy loop corrections.
A clue that this is the case comes from the consideration of phase transition contributions to vacuum energy in Sec. \ref{sec:phase}. There we have noted that while the phase transition contributions are small after the transition, they could end up dominating {\it before}. Slow-roll large field inflation can in fact be viewed as one such slow, second-order phase transition -- very slow, in fact, to ensure that at least ${\cal O}(60)$ efolds of accelerated expansion may occur.

To check this, the simplest way to proceed is to consider the field equations that follow from the action
\be
S= \int d^4 x \sqrt{g} \left[ \frac{M^2_{Pl}}{2} R  - \Lambda -  \frac{\lambda^4}{2} \left(\frac{(\partial \varphi)^2}{\lambda^2} + m^2 \varphi^2 \right)
 - {\lambda^4} {\cal L}(\lambda^{-2} g^{\mu\nu} , \Phi, \varphi) \right] +\sigma\left(\frac{ \Lambda}{\lambda^4 \mu^4}\right) 
 \, ,
 \label{actioninfl}
\ee
where $\varphi$ is the inflaton avatar and $\Phi$ are the fields that it decays into after the end of inflation, including the Standard Model. This guarantees that the vacuum energy corrections remain completely sequestered even if they involve inflaton loops. Further, the inflaton sector needs to have an internal mechanism which protects the flatness of its potential (in this case, the smallness of $m$ in (\ref{actioninfl})), and allows for sufficient reheating, as explained, for example, in \cite{flux}. We will ignore these -- important -- details here, and merely focus on showing that (\ref{actioninfl})
supports inflation driven by the quadratic potential of $\varphi$, which is sufficient to prove the consistency between vacuum sequestration and large field inflation. This means, we will focus only on the quadratic $\varphi$ potential in the matter sector. 

For our purposes, it suffices to show that there exists an inflating FRW flat (toroidal) cosmology driven by the field $\varphi$ in the slow roll, and including the historic average cosmological term. With this, after canonically normalizing $\varphi \to \varphi/\lambda$, $m \to m/\lambda $, the variational equations (\ref{varsl}), (\ref{eeqs}) reduce to
\be
3M_{Pl}^2 H^2 = \frac12 \dot \varphi^2 + \frac12 m^2 \varphi^2 + \Lambda_\textrm{eff} \, , ~~~  \Lambda_\textrm{eff} = - \frac{\int dt a^3 (\frac12 m^2 \varphi^2 - \frac14 \dot \varphi^2)}{\int dt a^3}\, , ~~~ \ddot \varphi + 3H \dot \varphi + m^2 \varphi = 0 \, .
\ee
In slow roll approximation, the field kinetic energy and acceleration terms are negligible, so this system simplifies to
\be
3M_{Pl}^2 H^2 = \frac12 m^2 \varphi^2 + \Lambda_\textrm{eff} \, , \qquad \Lambda_\textrm{eff} = - \frac{m^2}{2} \frac{\int dt a^3 \varphi^2 }{\int dt a^3}\, , \qquad
3H \dot \varphi + m^2 \varphi = 0 \, .
\ee
Now, it is clear that the integral for $\Lambda_\textrm{eff}$ picks up significant contributions only from the time interval during which $\varphi \ne 0$. In other words, $\int dt a^3 \varphi^2 \simeq \int_0^{t_{end}} dt a^3 \varphi^2$ where $t_{end}$ is the end of inflation. We can bound this by taking 
$\varphi \le {\rm few} \times M_{Pl}$ during the last 60 efolds, which implies that 
$\int dt a^3 \varphi^2 \la 100 M_{Pl}^2 \int_0^{t_{end}} dt a^3 \simeq 100 M_{Pl}^2 e^{3N}/H^4$ where $H$ is the Hubble scale during inflation, and $N$ the number of efolds. So we find that $|\Lambda_\textrm{eff}| \la 50 \frac{m^2 M^2_{Pl}}{H^4{\int dt a^3}} e^{3N}$.
But the integral in the denominator involves the full cosmic history, and so by Eq. (\ref{inttruni}) it is $\simeq 1/H_{age}^4$. This means,
\be
|\Lambda_\textrm{eff}| \la 50  \frac{m^2 M_{Pl}^2 H_{age}^4 e^{3N}}{H^4} \, .
\ee
So relative to the scale of inflation, this gives $\frac{|\Lambda_\textrm{eff}|}{M_{Pl}^2 H^2}  \la 50  \frac{m^2 H_{age}^4 e^{3N}}{H^6}$.
Now, to have slow roll inflation we must enforce $m<H$. More importantly, for inflation to solve the horizon problem, we must require that $H_0 e^N/H \le 1$. Since $H_{age} \le H_0$, this immediately shows that 
\be
\frac{|\Lambda_\textrm{eff}|}{M_{Pl}^2 H^2}  \ll 1 \, .
\ee
This means that the historic average contribution to the residual cosmological constant from the inflationary dynamics is completely negligible during inflation. Similar considerations also apply to other power law potentials. Hence large field inflation  proceeds as in GR. 

The one remaining point of potential conflict between our proposal and inflation is the regime of eternal inflation \cite{eternal}. For our proposal to work
we need the spacetime volume of the universe to be finite. On the other hand, in eternal inflation large quantum fluctuations of the inflaton field
restart inflation in various regions of space. In effect this allows the inflaton to utilize an entire ensemble of initial conditions, many of which yield neverending exponential expansion creating infinite spacetime volumes. Although the dynamics of eternal inflation remains a topic of debate, it is nevertheless an important ingredient of the inflationary paradigm and hence we cannot ignore it. 

The success of our mechanism in its present form precludes eternal inflation. We can imagine two ways to accomplish this. The simplest possibility is to  imagine that the true minima of the QFT have negative potential, in which case the vacuum would have been an anti-de Sitter. This is even 
consistent with the fact that our historic term $\langle \tau^\alpha{}_\alpha \rangle$ generically yields a very small {\it negative} residual cosmological constant. If this happens, then a random observer will eventually find herself in the global AdS minimum and experience an apocalyptic crunch. Of course, this does not rule out an infinite universe, as the birth-rate of inflating bubbles could exceed the death-rate due to crunches. Whether or not this is the case depends on the shape of the potential barrier separating inflationary vacua from the AdS minimum. A scenario in which the rate of terminal collapse of a universe exceeds their birth rate corresponds to the case where the barrier is neither too wide nor too steep \cite{sink,nima}. One could reasonably expect to achieve this and still allow the inflaton field to explore an ensemble of inflationary initial conditions. The convincing, quantitative estimates of the likelihood of such a scenario at this point are obstructed by the ambiguities in the definition of the cosmological measures, and dynamics of eternal inflation. Another possibility is to imagine that inflationary potential at high energies -- or large displacements of the inflaton from the vacuum -- is dramatically modified preventing inflationary slow roll altogether. For this, one needs to have very steep potentials away from the minimum. This does appear to be a designer model, but it is nevertheless possible to construct such theories \cite{sink,nima,lindenoneternal,matt}. In such setups, one can avoid eternal inflation but then one needs a different approach for addressing initial conditions for inflation.

\section{Summary} \label{sec:conc}

We have proposed a mechanism for stabilizing the cosmological constant from vacuum energy corrections. The main idea behind it is to modify the dynamics by introducing {\it two} global variables which yields a theory with two 
new approximate symmetries: global scale invariance of the matter sector, which is broken only by the Einstein-Hilbert term, and a shift symmetry that allows us to shift the matter Lagrangian by a constant without affecting the geometry. These symmetries provide the key insight into how the vacuum energy is sequestered. At a fixed scale below the cutoff, the shift symmetry is responsible for the cancellation of the vacuum energy after it is renormalized. The scaling symmetry then guarantees that the shift symmetry remains operational at all scales below the cutoff. Thus, all the vacuum energy corrections coming from a sector whose dynamics is constrained by the global variables is completely removed from the gravitational field equations. Since the two symmetries are approximate, being broken by the gravitational sector, the net residual cosmological constant is not zero, but it is automatically small in old large universes, being given by the historic average 
$\langle \tau^\alpha{}_\alpha \rangle= \int d^4x \sqrt{g} \tau^\alpha{}_\alpha/\int d^4 x \sqrt{g}$, where the trace involves only the contributions from the fluctuating on-shell sources that affect cosmic evolution. 

This term is non-local. However this non-locality is not pathological, but a mere consequence that the starting cosmological constant is a UV-divergent variable in the theory. It must be renormalized, which means that the finite leftover part is a quantity that cannot be predicted, but must be measured. Since the cosmological constant is a global variable, a parameter of a system of codimension zero, and its measurement requires carefully separating it from all the other long-wavelength modes in the universe, it takes a detector of the size of the universe to measure it. Once determined, however, it is independent of any vacuum energy corrections, and can be used to predict other, UV-insensitive, observable. Note that the cancellation of the UV-sensitive contributions is perfectly local in spacetime. 

This {\it aposteriority} of the measurement of $\Lambda_\textrm{eff}$ represents a self-consistent determination of a {\it single number}, and does not give rise to any pathologies one typically associates with local violations of causality. It is interesting to note Coleman has argued that  the universe should possess a degree of what he calls {\it precognition} -- it knows from the beginning that it intends to grow old and big \cite{cole1}.

All this follows from the modification of the gravitational sector, accomplished with the introduction of two global variables and two global constraints. This approach, equivalent to the isoperimetric problem of variational calculus, does not affect local particle physics at all. The theory still has diffeomorphism invariance and local Poincar\'e symmetry, and so local QFT behaves exactly the same as in the conventional formulation. Further, because of these symmetries the number of local degrees of freedom in gravity is still just two, the usual spin-2 helicities of GR. The global variables disappear from local dynamics, one canceling the quantum vacuum energy and the other being absorbed into the definition of physical scales of the local QFT, without affecting particle physics. A key ingredient of the proposal, however, which allows for this disappearance of the variable $\lambda$, is that the universe should be compact in space and time. This is necessary in the present formulation in order to 
to have nonzero mass gap in the QFT sector. So far we have not specified the detailed mechanism which produces collapse, but have shown that this is consistent with the dynamics of the proposal. In a forthcoming work \cite{KP3} we will return to this issue.
  
The main phenomenological differences of our proposal relative to the standard formulation of QFT coupled to gravity, and so predictions, follow from this requirement. This completely changes cosmic eschatology since we now require the universe to be spatially closed and to exist for a finite proper time, starting out with a Big Bang and ending with a Big Crunch. A spatially closed universe can have observational signatures, yielding doubling of images due to nontrivial spatial topology or affecting CMB at the largest scales. Currently there is no direct evidence of such topology \cite{topology}, but the search continues. Further, it may be possible to forecast the impending collapse observationally, as noted in \cite{doomsday}, particularly if the impending doom is  triggered by whatever is driving the current acceleration. The key consequence, however, is the prediction that the current epoch 
of cosmic acceleration with $w_{DE} \approx -1$ is a transient. So finding deviations away from this equation of state of dark energy may indeed be a  harbinger of a future cosmological collapse.

\section*{Acknowledgements}
We would like to thank Guido D'Amico, Ido Ben-Dayan, Ed Copeland, Savas Dimopoulos, Kiel Howe, Justin Khoury, Matt Kleban, Albion Lawrence, Andrei Linde, Adam Moss, Fernando Quevedo, Paul Saffin, Martin Sloth, Lorenzo Sorbo, David Stefanyszyn, Natalia Toro, Alex Vikman, Andrew Waldron and Alexander G. Westphal for useful discussions.  NK thanks the School of Physics and Astronomy, U. of Nottingham for hospitality in the course of this work.
NK is supported by the DOE Grant DE-FG03-91ER40674, and received support  from a Leverhulme visiting professorship.
AP was funded by a Royal Society URF.

\end{document}